\RequirePackage{fix-cm}
\documentclass[twocolumn, epjc3, comma, sort&compress, natbib]{svjour3}
\bibpunct{[}{]}{,}{n}{}{,} 

\journalname{Eur. Phys. J. A}

\usepackage{latexsym}
\usepackage{amsmath}
\usepackage{amssymb}
\usepackage{amsfonts}
\usepackage{CJKutf8}

\usepackage[mathscr,scaled=1.15]{urwchancal}
\DeclareFontFamily{OT1}{pzc}{}
\DeclareFontShape{OT1}{pzc}{m}{it}%
{<-> s * [1.15] pzcmi7t}{}
\DeclareMathAlphabet{\mathpzc}{OT1}{pzc}{m}{it}

\usepackage{color}

\usepackage{supertabular}
\usepackage{placeins}
\usepackage{epsfig}
\usepackage{graphicx}

\definecolor{purple}{rgb}{0.5,0,0.5}
\definecolor{blue}{rgb}{0.0,0,0.9}
\definecolor{prdblue}{rgb}{0.133,0.118,0.498}
\usepackage[colorlinks=true, pdfstartview=FitV, linkcolor=prdblue, citecolor= prdblue, urlcolor=prdblue]{hyperref}

\begin{document}
\begin{CJK*}{UTF8}{gbsn}

\title{$\,$\\[-6ex]\hspace*{\fill}{\normalsize{\sf\emph{Preprint no}.\
NJU-INP 101/25}}\\[1ex]
Distribution Functions of $\Lambda$ and $\Sigma^0$ Baryons}

\author{Yang Yu (俞杨)\thanksref{NJU,INP}%
        $\,^{\href{https://orcid.org/0009-0008-8011-3430}{\textcolor[rgb]{0.00,1.00,0.00}{\sf ID}}}$
\and
    Peng Cheng (程鹏)\thanksref{AHNU}%
    $\,^{\href{https://orcid.org/0000-0002-6410-9465}{\textcolor[rgb]{0.00,1.00,0.00}{\sf ID}}}$
\and
    Hui-Yu Xing (邢惠瑜)\thanksref{NJU,INP}%
    $\,^{\href{https://orcid.org/0000-0002-0719-7526}{\textcolor[rgb]{0.00,1.00,0.00}{\sf ID}}}$
\and
    Daniele Binosi\thanksref{ECT}%
    $\,^{\href{https://orcid.org/0000-0003-1742-4689}{\textcolor[rgb]{0.00,1.00,0.00}{\sf ID}}}$
\and
    Craig D.\ Roberts\thanksref{NJU,INP}%
       $\,^{\href{https://orcid.org/0000-0002-2937-1361}{\textcolor[rgb]{0.00,1.00,0.00}{\sf ID}}}$
}

\institute{School of Physics, \href{https://ror.org/01rxvg760}{Nanjing University}, Nanjing, Jiangsu 210093, China \label{NJU}
\and
Institute for Nonperturbative Physics, \href{https://ror.org/01rxvg760}{Nanjing University}, Nanjing, Jiangsu 210093, China
\label{INP}
\and
Department of Physics, \href{https://ror.org/05fsfvw79}{Anhui Normal University}, Wuhu, Anhui 24100, China\label{AHNU}
\and
European Centre for Theoretical Studies in Nuclear Physics
            and Related Areas  (\href{https://ror.org/01gzye136}{ECT*})\\ \hspace*{0.5em}Villa Tambosi, Strada delle Tabarelle 286, I-38123 Villazzano (TN), Italy\label{ECT}
%
           %
\\[1ex]
Email:
\href{mailto:cdroberts@nju.edu.cn}{cdroberts@nju.edu.cn} (CDR)
            }

\date{2025 July 29} 

\maketitle

\end{CJK*}

\begin{abstract}
Treating baryons as quark + interacting-diquark bound states, a symmetry-preserving formulation of a vector$\,\times\,$vector contact interaction (SCI) is used to deliver an extensive, coherent set of predictions for $\Lambda, \Sigma^0$ baryon unpolarised and polarised distribution functions (DFs) -- valence, glue, and four-flavour separated sea -- and compare them with those of a like-structured nucleon.  $\Lambda, \Sigma^0$ baryons are strangeness negative-one isospin partners within the SU$(3)$-flavour baryon octet.  This makes such structural comparisons significant.  The study reveals impacts of diquark correlations and SU$(3)$-flavour symmetry breaking on $\Lambda$, $\Sigma^0$ structure functions, some of which are significant.  For instance, were it not for the presence of axialvector diquarks in the $\Sigma^0$ at the hadron scale, the $s$ quark could carry none of the $\Sigma^0$ spin.  The discussion canvasses issues that include helicity retention in hard scattering processes; the sign and size of polarised gluon DFs; and the origin and decomposition of baryon spins.  Interpreted judiciously, the SCI analysis delivers an insightful explanation of baryon structure as expressed in DFs.
\end{abstract}

\section{Introduction}
Baryons are hadrons.
Each is seeded by three valence quarks.
This seems straightforward.
However, baryons are supposed to be described by quantum chromodynamics (QCD).
This changes the character of the problem because the path to a rigorous solution must pass nonperturbatively through the complexities of \linebreak Poincar\'e-invariant quantum non-Abelian gauge field theory.
In QCD, the degrees of freedom most appropriate for formulating a given problem depend on both the observable being addressed and the resolving scale of the probe being used.
These features and complexities make baryons the most fundamental and challenging three-body problems in Nature.

Baryon structural properties have long been the target of studies using numerical simulations of lattice-regularised QCD (lQCD), with a focus mostly on the proton; see, \emph{e.g}., Ref.\,\cite{Lin:2017snn}.  An alternative is provided by continuum Schwinger function methods (CSMs); see Ref.\,\cite{Roberts:1994dr} for an introduction.  Within this framework, the study of baryon bound states begins with a Poincar\'e-covariant Faddeev equation, the first direct solution of which was obtained in Ref.\,\cite{Eichmann:2009qa}.  There have been numerous subsequent applications \cite{Eichmann:2016yit}; and especially during the past decade, significant progress has been made.
For instance, today, a unified set of predictions is available for nucleon electromagnetic and gravitational form factors \cite{Yao:2024uej, Yao:2024ixu}.  Nevertheless, a direct solution of this QCD bound state problem remains challenging, demanding efficacious algorithms, precise numerical analysis, and high-performance computers.

\begin{figure}[t]
\centerline{%
\includegraphics[clip, height=0.14\textwidth, width=0.45\textwidth]{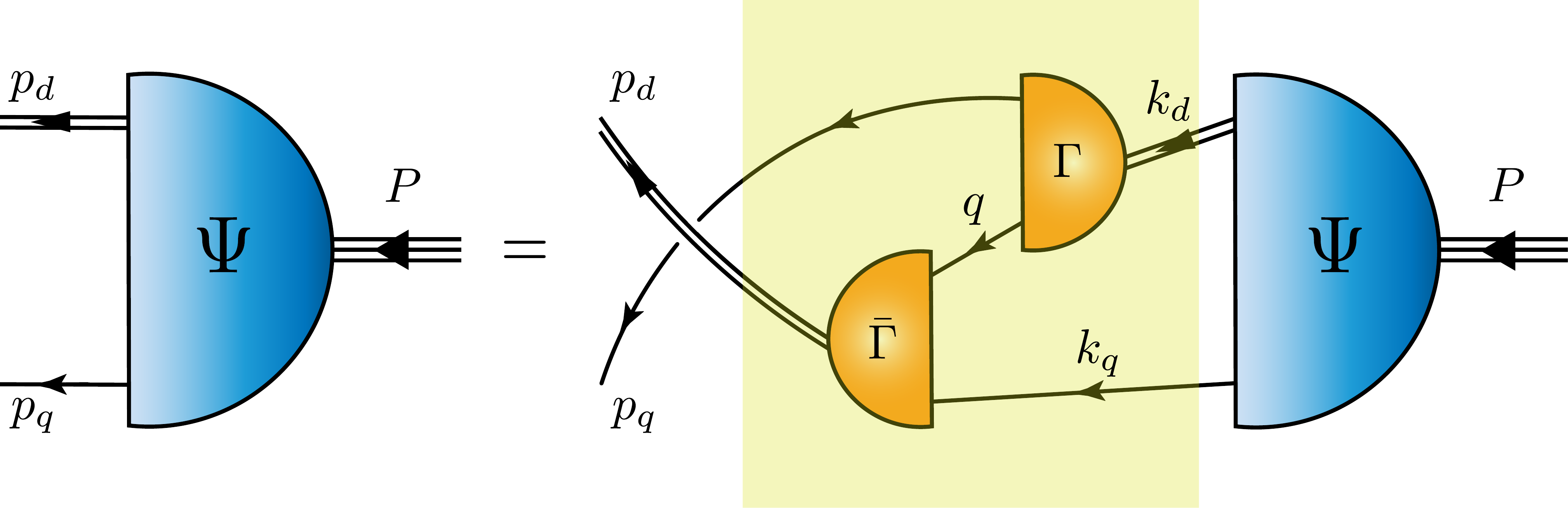}}
\caption{\label{FigFaddeev}
Homogeneous integral equation for the Poincar\'e-covariant matrix-valued function $\Psi$, which is the Faddeev amplitude for a baryon with total momentum $P=p_q+p_d=k_q+k_d$ built from three valence quarks, two of which are contained in a nonpointlike, fully-interacting diquark correlation. $\Psi$ describes the relative momentum correlation between the dressed-quarks and -diquarks.
Legend. \emph{Shaded rectangle} -- Faddeev kernel;
\emph{single line} -- dressed-quark propagator, Eq.\,\eqref{Squark};
$\Gamma$ -- diquark correlation amplitude, Eq.\,\eqref{qqBSAs}, and \emph{double line} -- diquark propagator, Eq.\,\eqref{qqPropagator}.
Only isoscalar--scalar and isovector--axialvector diquarks play a role in positive parity octet baryons \cite{Barabanov:2020jvn}.
Throughout, $[q_1q_2]$ denotes isoscalar--scalar diquark and $\{q_1q_2\}$ is isovector--axialvector.}
\end{figure}

A simpler version of the problem was introduced in Refs.\,\cite{Cahill:1988dx, Reinhardt:1989rw, Efimov:1990uz}, wherein by exploiting the pairing capacity of fermions, expressed in baryons by the formation of quark + quark (diquark) correlations \cite{Barabanov:2020jvn}, a quark + fully-interacting diquark Faddeev equation was derived; see Fig.\,\ref{FigFaddeev}.  Since the first, rudimentary numerical studies \cite{Burden:1988dt}, the approach has developed into a sophisticated tool that has been used to predict a large array of baryon observables \cite{Eichmann:2016yit, Burkert:2017djo, Brodsky:2020vco, Roberts:2020hiw, Carman:2023zke}.  Of particular relevance to the discussion herein are recent predictions for nucleon parton distribution functions (DFs) \cite{Lu:2022cjx, Cheng:2023kmt, Yu:2024qsd, Yu:2024ovn}.

Regarding the structure of the $J^P=(1/2)^+$ baryon octet, the modern quark + interacting-diquark picture predicts that isoscalar-scalar and isovector-axialvector diquarks must both be included in the Faddeev wave functions of these systems if a realistic description is to be obtained.
For the systems of primary interest herein, isoscalar--scalar diquarks are $\approx 0.26\,$GeV lighter than isovector--vector diquarks \cite{Yin:2019bxe, Yin:2021uom}.
Baryon structure models that ignore axialvector diquarks are incompatible with existing data; see, \emph{e.g}., Refs.\,\cite{Cui:2021gzg, Chen:2021guo, Chen:2023zhh}.

It is worth remarking that the picture is readily extended to SU$(5)$ flavour \cite{Yin:2019bxe, Yin:2021uom, Torcato:2023ijg}, \emph{i.e}., to systems involving one or more heavy quarks.  These analyses reveal that it is typically the lightest allowed diquark correlation that defines the most important component of a baryon’s
Faddeev amplitude.  For instance, the isoscalar--scalar $[uQ]$ diquark dominates in the $\Xi_{QQ}$, $Q=c,b$.  Furthermore, this remains true even if an isovector--axialvector diquark is the lightest channel \linebreak  available, \emph{e.g}. $\{uu\}$ is dominant in the $\Sigma_c$.
Such outcomes speak against the treatment of singly-heavy baryons as effectively two-body light-diquark + heavy-quark systems and doubly heavy baryons as two-body heavy-heavy diquark + light-quark systems.
Instead, deciding the dominant component of a baryon's Faddeev amplitude is a dynamical question, which is only answered by explicit case-by-case calculations.

Returning to DFs, predictions for those of the nucleon display clear signatures of diquark substructure.  However, little is known about their impact on the DFs of other baryons.  We remedy that herein by delivering predictions for DFs (valence, sea, glue) of the $\Lambda(1116)$ and $\Sigma^0(1193)$.  Contrasting these systems is of special interest because they have the same valence-quark content but different isospin: $\Lambda (I=0)$ and $\Sigma^0(I=1)$.  Consequently, in the quark + diquark picture, their spin-flavour wave functions have markedly different structure:
{\allowdisplaybreaks
\begin{subequations}
\label{FaddeevFlavour}
\begin{align}
u_\Lambda & =
\frac{1}{\sqrt 2}
\left[
\begin{array}{ll}
{\rm r}_1^0\,: & \sqrt{2}s[ud]           \\
{\rm r}_2^0\,: & d[us]-u[ds]     \\
{\rm r}_3^0\,: & d\{us\}-u\{ds\}
\end{array}\label{FALambdaA}
\right] ,\\
u_{\Sigma^0} & =
\frac{1}{\sqrt 2}
\left[
\begin{array}{ll}
{\rm r}_1^1\,: & d [us] + u [ds]           \\
{\rm r}_2^1\,: & 2 s\{ud\}     \\
{\rm r}_3^1\,: & d\{us\} + u\{ds\}
\end{array}
\right] . \label{FaddeevFlavourS}
\end{align}
\end{subequations}
The $\Lambda$ has two isoscalar-scalar components and one isovector-vector, whereas the $\Sigma^0$ has one isoscalar-scalar contribution and two isovector-vector.
One should expect expressions of these differences in the unpolarised and helicity dependent DFs of these systems.
}

Throughout, we will treat $u, d$ quarks as mass-de\-ge\-ne\-rate; consequently, all results will exhibit isospin symmetry.
Notably, isospin symmetry is not the same as SU$(3)$-flavour symmetry; so, for instance, $\Sigma^0$ and $\Lambda$ are not mass-degenerate.

For reference, we list the neutron spin-flavour structure:
\begin{align}
u_n & =
\left[
\begin{array}{ll}
{\rm r}_1^n\,: & d[ud]           \\
{\rm r}_2^n\,: & u\{dd\}     \\
{\rm r}_3^n\,: & d\{ud\}
\end{array}
\right]\,.
\label{protonSF}
\end{align}
The proton, $p$, is obtained under $u\leftrightarrow d$ interchange.

Studies of the DFs of octet baryons with strangeness have been completed using an array of tools; see, \emph{e.g}., Refs.\,\cite{Boros:1999da, Ma:2001ri, Gockeler:2002uh, Erkol:2009ek, CSSM:2014uyt, Peng:2022lte}.
Herein, we approach the problem using a symmetry preserving treatment of a vector\,$\times$\,vector contact interaction (SCI) \cite{GutierrezGuerrero:2010md, Roberts:2010rn, Roberts:2011wy, Wilson:2011aa}.
The SCI has been used fruitfully in a wide variety of studies, with those of most relevance to our analysis being focused on baryon spectra and interactions \cite{Yin:2019bxe, Yin:2021uom, Cheng:2022jxe, Yu:2024qsd}.

The SCI is not a precision tool, but it does have many merits, \emph{e.g}.:
algebraic simplicity;
simultaneous applicability to a diverse array of systems and processes;
potential for revealing insights that link and explain numerous phenomena;
and capability to serve as a tool for checking the validity of algorithms employed in calculations that depend upon high performance computing.
Significantly, today's applications are typically parameter-free.
Brief recapitulations of SCI features and formulation are provided in the appendices of Refs.\,\cite{Yin:2019bxe, Yin:2021uom, Cheng:2022jxe, Yu:2024qsd}.
When necessary, relevant results are reproduced herein.

Our presentation is parcelled as follows.
Section~\ref{SolveFE} reviews relevant aspects of the SCI treatment of baryons as a quark + diquark problem and reports solutions for $\Lambda, \Sigma^0, p$.
Algebraic SCI formulae for helicity-independent $\Lambda, \Sigma^0$ DFs are developed and explained in Sect.\,\ref{HSHIDFs}.
In Sect.\,\ref{HSHDDFs}, this is repeated for the helicity-dependent DFs.
Numerical results for all these DFs at the hadron scale, $\zeta_{\cal H}$, are presented and discussed in Sect.\,\ref{DFresultsHS}.  The hadron scale is uniquely defined as that resolving scale whereat all properties of a given hadron are carried by its valence degrees of freedom \cite{Yin:2023dbw}.
Scale evolution \cite{Dokshitzer:1977sg, Gribov:1971zn, Lipatov:1974qm, Altarelli:1977zs} of hadron DFs is discussed in Sect.\,\ref{EvolvedDFs}.  We use the all-orders (AO) scheme, which is based on the notion of effective charges \cite{Grunberg:1980ja, Grunberg:1982fw, Deur:2023dzc} and is detailed in Ref.\,\cite{Yin:2023dbw}.
Implications of our results for issues relating to baryon spin decompositions are discussed in Sect.\,\ref{SecSpin}.
Section~\ref{epilogue} is a summary and perspective.

\section{Solutions of the SCI Faddeev Equation}
\label{SolveFE}
To obtain the $\Lambda, \Sigma^0$ Faddeev amplitudes, one must solve the equation defined implicitly by Fig.\,\ref{FigFaddeev}.
Its kernel involves the SCI propagator for a dressed quark:
\begin{equation}
S_f(k) = 1/[i\gamma\cdot k + M_f]\,,
\label{Squark}
\end{equation}
where $f=u,d,s$ labels the quark flavour and $M_f$ is the associated dressed mass;
the correlation amplitudes of $[ud]$, $[(l=u=d)s]$, $\{ud\}$, $\{ls\}$ diquarks,
\begin{subequations}
\label{qqBSAs}
\begin{align}
\Gamma^{[fg]}(K) & =
\gamma_5\left[ i E_{[fg]} + \frac{\gamma\cdot K}{2 M_{fg}} F_{[fg]} \right]C,\\
\Gamma_\mu^{\{fg\}} & = T_{\mu\nu}^K \gamma_\nu C E_{\{fg\}}\,,
\end{align}
\end{subequations}
$f,g= u,d,s$, where $K$ is the correlation's total momentum, $C=\gamma_2\gamma_4$ is the charge conjugation matrix, $M_{fg} = M_f M_g/[M_f + M_g]$;
and associated propagators for these correlations,
\begin{subequations}
\label{qqPropagator}
\begin{align}
\Delta^{[fg]}(K) & = \frac{1}{K^2 + m_{[fg]}^2}\,, \\
\Delta^{\{fg\}}_{\mu\nu}(K) & =\delta_{\mu\nu} \frac{1}{K^2 + m_{\{fg\}}^2}\,,
\label{AVdqprop}
\end{align}
\end{subequations}
where $m_{[fg]}$, $m_{\{fg\}}$ are the correlations' masses.

\begin{table}[t]
\caption{\label{Tab:DressedQuarks}
Couplings, $\alpha_{\rm IR}/\pi$, ultraviolet cutoffs, $\Lambda_{\rm uv}$, and current-quark masses, $m_f$, $f=u/d,s$, that deliver a good description of flavoured pseudoscalar meson properties, along with the dressed-quark masses, $M_f$, pseudoscalar meson masses, $m_{P}$, and leptonic decay constants, $f_{P}$, they produce; all obtained with $m_G=0.5\,$GeV, $\Lambda_{\rm ir} = 0.24\,$GeV.
Empirically, at a sensible level of precision \cite{ParticleDataGroup:2024cfk}:
$m_\pi =0.14$, $f_\pi=0.092$;
$m_K=0.50$, $f_K=0.11$.
%
(We assume isospin symmetry and list dimensioned quantities in GeV.)}
\begin{center}
\begin{tabular*}
{\hsize}
{
l@{\extracolsep{0ptplus1fil}}|
c@{\extracolsep{0ptplus1fil}}|
c@{\extracolsep{0ptplus1fil}}
c@{\extracolsep{0ptplus1fil}}
c@{\extracolsep{0ptplus1fil}}
|c@{\extracolsep{0ptplus1fil}}
c@{\extracolsep{0ptplus1fil}}
c@{\extracolsep{0ptplus1fil}}}\hline
& quark & $\alpha_{\rm IR}/\pi\ $ & $\Lambda_{\rm uv}$ & $m_{\rm current}$ &   $M_{\rm dressed}$ &  $m_{P}$ & $f_{P}$ \\\hline
$\pi\ $  & $l=u/d\ $  & $0.36\phantom{2}$ & $0.91\ $ & $0.0068_{u/d}\ $ & 0.37$\ $ & 0.14 & 0.10  \\\hline
$K\ $ & $\bar s$  & $0.33\phantom{2}$ & $0.94\ $ & $0.16_s\phantom{7777}\ $ & 0.53$\ $ & 0.50 & 0.11 \\\hline
\end{tabular*}
\end{center}
\end{table}

Each of the quantities required to complete the expressions in Eqs.\,\eqref{Squark}\,--\,\eqref{qqPropagator} was calculated in Ref.\,\cite{Cheng:2022jxe}.
It is worth highlighting that they are not independent.  Instead, every one is completely determined after the SCI interaction strength and the $l$, $s$ current masses are fixed.
The SCI formulation parameters are listed in Table~\ref{Tab:DressedQuarks} along with the dressed quark masses, pseudoscalar meson masses and decay constants they produce.
The calculated diquark masses and canonically normalised amplitudes determined therewith are listed in Table~\ref{qqBSAsolutions}.

\begin{table}[t]
\caption{\label{qqBSAsolutions}
Masses and canonically normalised correlation amplitudes obtained by solving the diquark Bethe-Salpeter equations.
(Isospin symmetry is assumed.  Masses are listed in GeV.  Amplitudes are dimensionless.)}
\begin{center}
\begin{tabular*}
{\hsize}
{
c@{\extracolsep{0ptplus1fil}}
c@{\extracolsep{0ptplus1fil}}
c@{\extracolsep{0ptplus1fil}}|
c@{\extracolsep{0ptplus1fil}}
c@{\extracolsep{0ptplus1fil}}
c@{\extracolsep{0ptplus1fil}}}\hline
$m_{[ud]}$ & $E_{[ud]}$ & $F_{[ud]}\ $ & $m_{[ls]}$ & $E_{[ls]}$ & $F_{[ls]}\ $ \\
$0.78$ & $2.71$ &  $0.31\ $ &  $0.94$  & $2.78$ & $0.37\ $\\
$m_{\{ud\}}$ & $E_{\{ud\}}$ & & $m_{\{ls\}}$ & $E_{\{ls\}}$ &  \\
$1.06$ & $1.39$ &  &  $1.22$  & $1.16$ & \\\hline
\end{tabular*}
\end{center}
\end{table}

The solution of the SCI Faddeev equation can be written in the form $\Psi_B(P) = \psi_B(P) u(P)$,
where $B$ labels the baryon;
$u(P)$ is a positive energy Dirac spinor, whose properties are detailed in Ref.\,\cite[Appendix A.3]{Cheng:2022jxe}; and
\begin{equation}
\label{FALambda}
\psi_I(P) u(P) = \sum_{d=r_{1,2,3}^I} c_{d} \psi^{d}(P) u(P)\,,
\end{equation}
with $\Lambda(I=0)$, $\Sigma^0(I=1)$.
In common with typical SCI formulations, the Faddeev amplitude herein does not depend on the quark-diquark relative momentum.
Generalising the scheme to include such dependence only brings complexity without additional insight \cite{Xu:2015kta}.

{\allowdisplaybreaks
To complete the specification of the flavour-spin \linebreak structure of the SCI baryon amplitudes in Eq.\,\eqref{FALambda}, we record the following:
\begin{subequations}
\begin{align}
c^{r_1^0} \psi_{r_1^0} & = {\mathpzc s}^{r_1^0} I_{\rm D} \,, \label{cr10}\\
c^{r_2^0} \psi_{r_2^0} & = {\mathpzc s}^{r_2^0} I_{\rm D} \,, \label{cr20}\\
c^{r_3^0} \psi_\mu^{r_3^0} & =
{\mathpzc a}_1^{r_3^0} i \gamma_5\gamma_\mu
+ {\mathpzc a}_2^{r_3^0} \gamma_5 \hat P_\mu \,, \label{cr30} \\
c^{r_1^1} \psi_{r_1^1} & = {\mathpzc s}^{r_1^1} I_{\rm D}\,,  \\
c^{r_2^1} \psi_\mu^{r_2^1} & = {\mathpzc a}_1^{r_2^1} i \gamma_5\gamma_\mu
+ {\mathpzc a}_2^{r_2^1} \gamma_5 \hat P_\mu\,, \label{cr21}\\
c^{r_3^1} \psi_\mu^{r_3^1} & = {\mathpzc a}_1^{r_3^1} i \gamma_5\gamma_\mu
+ {\mathpzc a}_2^{r_3^1} \gamma_5 \hat P_\mu \,, \label{cr31}
\end{align}
\end{subequations}
where $\hat P^2=-1$.  With these things established and employing the procedure explained in Ref.\,\cite[Appendix A.3]{Cheng:2022jxe}, the solutions of the SCI Faddeev equation for the $\Lambda$, $\Sigma^0$ baryons are readily obtained.  The masses and solution coefficients are listed in Table~\ref{FadSolution}.
}

\begin{table}[t]
\caption{\label{FadSolution}
Unit normalised Faddeev equation solutions relevant herein.
Empirically, the masses of the $\Lambda$, $\Sigma^0$ baryons are: $m_\Lambda \approx 1.12\,$GeV and $m_{\Sigma^0} \approx 1.19\,$GeV, with $m_{\Sigma^0}-m_\Lambda \approx 0.077\,$GeV.
Canonically normalised amplitudes are obtained by dividing the amplitude entries in each column by the following numbers:
${\mathpzc n}_c^{\Lambda}=0.217 = {\mathpzc n}_c^{\Sigma}$.
(The analogous factor for the nucleon amplitude is ${\mathpzc n}_c^{p,n} = 0.287$.)
}
\begin{center}
\begin{tabular*}
{\hsize}
{
l@{\extracolsep{0ptplus1fil}}|
c@{\extracolsep{0ptplus1fil}}|
c@{\extracolsep{0ptplus1fil}}}\hline
 & $\Lambda (I=0)\ $ & $\Sigma^0 (I=1)\ $ \\ \hline
mass/GeV$\ $  & $1.33\ $ & $1.38\ $ \\
${\mathpzc s}^{r_1^I}\ $ & $\phantom{-}0.66\phantom{0}\ $ & $\phantom{-}0.85\phantom{1}\ $ \\
${\mathpzc s}^{r_2^I}\ $ & $\phantom{-}0.62\phantom{0}\ $ & \\
${\mathpzc a}_1^{r_2^I}\ $ & & $-0.46\phantom{1}\ $ \\
${\mathpzc a}_2^{r_2^I}\ $ & & $\phantom{-}0.15\phantom{1}\ $ \\
${\mathpzc a}_1^{r_3^I}\ $ & $-0.41\phantom{0}\ $ & $\phantom{-}0.22\phantom{1}\ $ \\
${\mathpzc a}_2^{r_3^I}\ $ & $-0.084\ $ & $\phantom{-}0.041\ $ \\
\hline
\end{tabular*}
\end{center}
\end{table}

Following a long established pattern \cite{Roberts:2011cf}, the masses in Table~\ref{FadSolution} are deliberately $\approx 0.20\,$GeV above experiment \cite{ParticleDataGroup:2024cfk} because Fig.\,\ref{FigFaddeev} describes the \emph{dressed-quark core} of each baryon.  To build a complete baryon, resonant contributions should also be included in the Faddeev kernel.  Such ``meson cloud'' dynamics is known to lower the mass of octet baryons by $\approx 0.2$\,GeV \cite{Roberts:2011cf, Hecht:2002ej, Sanchis-Alepuz:2014wea}.  (Likewise, in quark models \cite{Garcia-Tecocoatzi:2016rcj, Chen:2017mug}.)  Whilst their impact on baryon structure can be estimated using dynamical coupled-channels models \cite{Aznauryan:2012ba, Burkert:2017djo}, that is beyond the scope of contemporary Faddeev equation analyses.  Notwithstanding these things, the quark + dynamical diquark picture readily produces a result for $m_{\Sigma^0}-m_\Lambda$ that is commensurate with experiment.
This is because the $I=0$ $\Lambda$-baryon is principally a scalar diquark system, whereas the $I=1$ $\Sigma^0$ has more axialvector strength: scalar diquarks are lighter than axialvector diquarks; see Table~\ref{qqBSAsolutions}.

As already noted, the Faddeev amplitudes in Table~\ref{FadSolution} are unit normalised.
However, one must use the canonically normalised amplitude in the calculation of observables.
That is defined via the baryon's elastic electromagnetic Dirac form factors, $F_1^{Bq}(Q^2=0)$, where $q$ ranges over each valence quark in the state $B$.
It is straightforward to compute the single constant factor that, when used to rescale the unit-normalised Faddeev amplitude for $B$, ensures $F_1^{Bq}(0)=1$ $\forall q \in B$.
So long as one employs a symmetry-preserving treatment of the elastic electromagnetic scattering problem, it is certain that a single factor ensures all such flavour-separated form factors are unity at $Q^2=0$.
Explicit examples are provided elsewhere \cite{Wilson:2011aa}.

\section{Hadron-Scale Helicity-Independent Parton Distribution Functions -- Algebraic Formulae}
\label{HSHIDFs}
Following Ref.\,\cite{Cui:2020tdf}, it has become apparent that a unique definition of a hadron scale, $\zeta_{\cal H}$, is possible.
Namely, it is the probe scale whereat valence quasiparticle degrees-of-freedom should be used to formulate and solve hadron bound state problems.
The notion is proving efficacious -- see, \textit{e.g}., Refs.\,\cite{Ding:2022ows, Carman:2023zke, Raya:2024ejx, Yao:2024ixu, Xu:2024nzp} -- and is a basic tenet of the AO DF evolution scheme \cite{Yin:2023dbw}.
We exploit these ideas herein and discuss them further below.

The question of a role for an explicit Wilson line in the definition and calculation of DFs might be raised in this connection.
In our approach, gluon (and sea quark) parton contributions are present at $\zeta_{\cal H}$: they are sublimated into the valence degrees-of-freedom -- so, hidden -- and exposed subsequently by AO evolution, which``undresses'' the quasiparticles from which the hadrons are built at $\zeta_{\cal H}$.
Practically, where valid comparisons are possible, there is agreement between DFs obtained in this way and those obtained using the manifestly gauge invariant lQCD approach to DF computation; see, \emph{e.g}., Refs.\,\cite{Lu:2022cjx, Lu:2023yna, Xu:2024nzp, Alexandrou:2024zvn}.
This can be interpreted as empirical justification for the approach employed herein.
Nevertheless, exploring the role that might be played by an explicit Wilson line is worth mathematical examination elsewhere.

It is also worth observing that owing to isospin symmetry, in both $\Lambda, \Sigma^0$, the $u$ and $d$ quark unpolarised DFs are separately identical.
So, we only calculate and describe $l=u=d$, $s$ DFs in these states.
Furthermore, since quark exchange diagrams are absent from the SCI interaction current, the DFs do not exhibit sensitivity to the relative signs between diquark factors in Eqs.\,\eqref{FaddeevFlavour}.

\subsection{$\Lambda$ baryon -- helicity independent}
Referring to Eq.\,\eqref{FaddeevFlavour} and Table~\ref{FadSolution}, one sees that the dominant configuration in the $\Lambda$ is $s[ud]$; hence, we begin with the hadron-scale valence $s$-in-$\Lambda$ DF.
Reviewing and adapting the proton analysis in Ref.\,\cite{Yu:2024qsd}, it becomes apparent that this DF receives five independent contributions:
\begin{align}
{\mathpzc s}_V^{\Lambda}(x;\zeta_{\cal H}) =
\sum_{t=Q^{[ud]}, D^{[us]}, D^{ [ds]}, D^{\{us\}}, D^{\{ds\}}} {\mathpzc s}_{V_t}^{\Lambda}(x;\zeta_{\cal H})\,.
\label{sDF(Lambda) eq(2)}
\end{align}

The first term describes the probe striking an $s$-quark that is accompanied by a bystander $[ud]$ diquark ($\hat\delta_n^{xP} = n\cdot P \delta(n\cdot \ell - x n\cdot P)$, $n^2=0$, $n\cdot P=-m_{\Lambda}$ in the $\Lambda$ rest frame):
\begin{align}
\Lambda_+ & \gamma\cdot n  {\mathpzc s}_{V_{Q^{[ud]}}}^{\Lambda}(x;\zeta_{\cal H}) \Lambda_+   =
\int\! \tfrac{d^4\ell}{(2\pi)^4} \,
\hat\delta_n^{xP} \;
\Lambda_+ Q^{[ud]}  \Lambda_+\,, \label{sDF1(Lambda) eq(3)}
\end{align}
with
$\Lambda_+ =(m_{\Lambda}-i\gamma \cdot P)/(2m_{\Lambda})$, $P^2=-m_{\Lambda}^2$; and
\begin{align}
Q^{[ud]} & = [c^{r_1^0}]^2\,
\bar\psi^{r_1^0}(-P) S_{s}(\ell)  \nonumber \\
& \quad \times \gamma\cdot n S_{s}(\ell) \psi^{r_1^0}(P) \Delta^{[ud]}(\ell-P)\,,
\label{Q^[ud]_0 eq(4)}
\end{align}
where Eq.\,\eqref{cr10} is implicit and, with $(\cdot)^{\rm T}$ denoting matrix transpose, one has the SCI identity
\begin{equation}
\bar\psi(P) = C^\dagger  \psi(P)^{\rm T} C = \psi(P) \,.
\end{equation}

The next two terms expose the $s$-quark within the $[ls]$ diquarks.
They can be expressed as convolutions:
\begin{align}
{\mathpzc s}_{V_{D^{[ls]}}}^{\Lambda}(x;\zeta_{\cal H})
&= \int_x^1\,\frac{dy}{y}\,{\mathpzc f}_{D^{[ls]}}^{\Lambda}(y;\zeta_{\cal H})
{\mathpzc s}_V^{[ls]}(x/y;\zeta_{\cal H})\,.
\label{sDF2and3(Lambda) eq(5)}
\end{align}
Here, ${\mathpzc s}_V^{[ls]}(x;\zeta_{\cal H})$ is the valence $s$-quark DF in an $[ls]$ diquark, discussed in \ref{diquarkDFs};
and, using the SCI, the probability density for finding an $[ls]$ diquark carrying a light-front fraction $x$ of the $\Lambda$-baryon's momentum is
\begin{subequations}
\label{fDls}
\begin{align}
\Lambda_+ \gamma & \cdot n  {\mathpzc f}_{D^{[ls]}}^{\Lambda}(x;\zeta_{\cal H}) \Lambda_+   =
\int\! \tfrac{d^4\ell}{(2\pi)^4} \,
\hat\delta_n^{xP} \;
\bar\Lambda_+ D^{[ls]}  \bar\Lambda_+\,,\label{f_{D^[qs]_0} eq(6a)} \\
D^{[ls]} & =  [c^{r_2^0}]^2\,
\bar\psi^{r_2^0}(P) S_l(\ell-P) \nonumber \\
& \qquad  \times \psi^{^{r_2^0}}(-P) i n\cdot \partial^\ell \Delta^{[ls]}(\ell)\, , \label{D^[qs]_0 eq(6b)}
\end{align}
\end{subequations}
where Eq.\,\eqref{cr20} is implicit and $\bar\Lambda_+ = (m_{\Lambda}+i\gamma \cdot P)/(2m_{\Lambda})$.

The final two terms expose the $s$-quark within $\{ls\}$ diquarks.
They can be also expressed as convolutions:
\begin{align}
{\mathpzc s}_{V_{D^{\{ls\}}}}^{\Lambda}(x;\zeta_{\cal H})
&= \int_x^1\,\frac{dy}{y}\,{\mathpzc f}_{D^{\{ls\}}}^{\Lambda}(y;\zeta_{\cal H})
{\mathpzc s}_V^{{\{ls\}}}(x/y;\zeta_{\cal H})\,.
\label{sDF4and5(Lambda) eq(7)}
\end{align}
Here, ${\mathpzc s}_V^{{\{ls\}}}(x;\zeta_{\cal H})$ is the valence $s$-quark DF in an $\{ls\}$ diquark, \ref{diquarkDFs}; and the probability density for finding an $\{ls\}$ diquark carrying a light-front fraction $x$ of the $\Lambda$-ba\-ryon's momentum is
\begin{subequations}
\label{fDlsav}
\begin{align}
\Lambda_+ \gamma \cdot & n {\mathpzc f}_{{D^{\{ls\}}}}^{\Lambda}(x;\zeta_{\cal H}) \Lambda_+
= \int\! \tfrac{d^4\ell}{(2\pi)^4} \, \hat\delta_n^{xP}\;
\bar \Lambda_+ D^{\{ls\}}  \bar \Lambda_+\,,\label{f_{D^{qs}_1} eq(8a)}\\
D^{\{ls\}} & =   c^{{r_3^0}}
\bar\psi_\rho^{{r_3^0}}(P)  S_l(\ell-P)   \nonumber \\
 &\qquad \times
c^{{r_3^0}} \psi_\sigma^{{r_3^0}}(-P)
i n\cdot \partial^\ell\Delta_{\rho\sigma}^{\{ls\}}(\ell) \,\label{D^{qs}_1 eq(8b)},
\end{align}
\end{subequations}
where Eq.\,\eqref{cr30} is implicit, which here means that this expression yields four distinct terms.

Turning now to the hadron-scale valence $u$-in-$\Lambda$ DF, there are again five terms:
\begin{align}
{\mathpzc u}_V^{\Lambda}(x;\zeta_{\cal H}) & =
\rule{-1em}{0ex}  \sum_{t=Q^{ [ds]}, Q^{\{ds\}}, D^{[ud]}, D^{[us]}, D^{\{us\}}}
\rule{-1em}{0ex} {\mathpzc u}_{V_t}^{\Lambda}(x;\zeta_{\cal H})\,.
\label{uDF(Lambda) eq(9)}
\end{align}
The first term in the sum counts the probe striking a $u$-quark in the presence of a bystander $[ds]$ diquark:
\begin{subequations}
\label{T1QuLambda}
\begin{align}
\Lambda_+ & \gamma\cdot n  {\mathpzc u}_{V_{Q^{[ds]}}}^{\Lambda}(x;\zeta_{\cal H}) \Lambda_+   =
\int\! \tfrac{d^4\ell}{(2\pi)^4} \,
\hat\delta_n^{xP} \;
\Lambda_+ Q^{[ds]}  \Lambda_+\,, \label{uDF1(Lambda) eq(10)} \\
Q^{[ds]} & = [c^{r_2^0}]^2 \,
\bar\psi^{r_2^0}(-P) S_l(\ell)  \nonumber \\
& \qquad \times  \gamma\cdot n S_l(\ell) \psi^{r_2^0}(P) \Delta^{[ls]}(\ell-P)\,,
\label{Q^[qs]_0 eq(11)}
\end{align}
\end{subequations}
with Eq.\,\eqref{cr20} implicit.

Term 2 in Eq.\,\eqref{uDF(Lambda) eq(9)} adds the probe striking a $u$-quark when the bystander is a $\{ds\}$ diquark:
\begin{subequations}
\label{T2QuLambda}
\begin{align}
\Lambda_+ & \gamma\cdot n  {\mathpzc u}_{V_{Q^{\{ds\}}}}^{\Lambda}(x;\zeta_{\cal H}) \Lambda_+   =
\int\! \tfrac{d^4\ell}{(2\pi)^4} \,
\hat\delta_n^{xP} \;
\Lambda_+ Q^{\{ds\}}  \Lambda_+ , \label{uDF2(Lambda) eq(12)} \\
Q^ {\{ds\}} & = c^{r_3^0}
\bar\psi_\rho^{r_3^0}(-P) S_l(\ell)  \nonumber \\
& \qquad \times \gamma\cdot n S_l(\ell) c^{r_3^0}\psi_\sigma^{r_3^0}(P) \Delta_{\rho\sigma}^{\{ls\}}(\ell-P)\,\label{Q^{qs}_1 eq(13)}.
\end{align}
\end{subequations}
Using Eq.\,\eqref{cr30}, this expands to four terms.

The third term on the right-hand side of Eq.\,\eqref{uDF(Lambda) eq(9)} catches the $u$-quark within the $[ud]$ diquark; and as in the earlier cases, it is expressed as a convolution:
\begin{align}
{\mathpzc u}_{V_{D^{[ud]}}}^{\Lambda}(x;\zeta_{\cal H}) = \int_x^1\,\frac{dy}{y}\,{\mathpzc f}_{D^{[ud]}}^{\Lambda}(y;\zeta_{\cal H})
{\mathpzc u}_V^{{[ud]}}(x/y;\zeta_{\cal H})\,.
\label{uDF3(Lambda) eq(14)}
\end{align}
Here, ${\mathpzc u}_V^{{[ud]}}(x;\zeta_{\cal H})$
is the valence $u$-quark DF in the $[ud]$ diquark, \ref{diquarkDFs}; and the SCI probability density for finding a $[ud]$ diquark carrying a light-front fraction $x$ of the $\Lambda$'s momentum is
\begin{subequations}
\begin{align}
\Lambda_+ \gamma & \cdot n  {\mathpzc f}_{D^{[ud]}}^{\Lambda}(x;\zeta_{\cal H}) \Lambda_+   =
\int\! \tfrac{d^4\ell}{(2\pi)^4} \,
\hat\delta_n^{xP} \;
\bar\Lambda_+ D^{[ud]}  \bar\Lambda_+\,,\label{f_{D^[ud]_0} eq(15a)}\\
D^{[ud]} & =  [c_{r_1^0}]^2
\bar\psi^{r_1^0}(P)  S_{s}(\ell-P) \nonumber \\
& \qquad \times  \psi^{r_1^0}(-P)
i n\cdot \partial^\ell \Delta^{[ud]}(\ell) \,,\label{D^[ud]_0 eq(15b)}
\end{align}
\end{subequations}
where Eq.\,\eqref{cr10} is implicit.

Term $4$ on the right-hand side of Eq.\,\eqref{uDF(Lambda) eq(9)} incorporates the DF contribution from the $u$-quark within the $[us]$ diquark:
\begin{align}
{\mathpzc u}_{V_{D^{[us]}}}^{\Lambda}(x;\zeta_{\cal H}) = \int_x^1\,\frac{dy}{y}\,
{\mathpzc f}_{D^{[ls]}}^{\Lambda}(y;\zeta_{\cal H})
{\mathpzc l}_V^{[ls]}(x/y;\zeta_{\cal H})\,.
\label{uDF4(Lambda) eq(16)}
\end{align}
Here, ${\mathpzc l}_V^{{[ls]}}(x;\zeta_{\cal H})$
is the valence $l$-quark DF in the $[ls]$ diquark, \ref{diquarkDFs}, and
${\mathpzc f}_{D^{[ls]}}^{\Lambda}(x;\zeta_{\cal H})$ is given in Eq.\,\eqref{fDls}.

The final term on the right-hand side of Eq.\,\eqref{uDF(Lambda) eq(9)} counts the $u$-quark within the $\{us\}$ diquark:
\begin{align}
{\mathpzc u}_{V_{D^{\{us\}}}}^{\Lambda}(x;\zeta_{\cal H}) = \int_x^1\,\frac{dy}{y}\,
{\mathpzc f}_{D^{\{ls\}}}^{\Lambda}(y;\zeta_{\cal H})
{\mathpzc l}_V^{{\{ls\}}}(x/y;\zeta_{\cal H})\,,
\label{uDF5(Lambda) eq(17)}
\end{align}
where ${\mathpzc l}_V^{{\{ls\}}}(x;\zeta_{\cal H})$
is the valence $l$-quark DF in an $\{ls\}$ diquark, \ref{diquarkDFs}, and
${\mathpzc f}_{D^{\{ls\}}}^{\Lambda}(x;\zeta_{\cal H})$ is given in \linebreak Eq.\,\eqref{fDlsav}.

Using elementary algebra, one can readily establish the following identities:
\begin{subequations}
\label{ExchangeLambda}
\begin{align}
{\mathpzc s}_{V_{Q^{[ud]}}}^{\Lambda}(x;\zeta_{\cal H})
 &= {\mathpzc f}_{D^{[ud]}}^{\Lambda}(1-x;\zeta_{\cal H}), \label{identity1 eq(19)}\\
{\mathpzc u}_{V_{Q^{[ds]}}}^{\Lambda}(x;\zeta_{\cal H})
&= {\mathpzc f}_{D^{[ls]}}^{\Lambda}(1-x;\zeta_{\cal H}), \label{identity2 eq(20)}\\
{\mathpzc u}_{V_{Q^{\{ds\}}}}^{\Lambda}(x;\zeta_{\cal H}) &= {\mathpzc f}_{D^{\{ls\}}}^{\Lambda}(1-x;\zeta_{\cal H}) \label{identity3 eq(21)}.
\end{align}
\end{subequations}
In concert with canonical normalisation of the $\Lambda$ Faddeev amplitude, they guarantee conservation of baryon number and momentum:
\begin{subequations}
\label{SumRules(Lambda) eq(22)}
\begin{align}
\int_0^1 dx\, & {\mathpzc l}_V^{\Lambda}(x;\zeta_{\cal H})  = 1\,,\;
%
%
\int_0^1 dx\, {\mathpzc s}_V^{\Lambda}(x;\zeta_{\cal H})  = 1\,, \label{baryonnumber(Lambda) eq(22a)}\\
& 2 \langle x \rangle_{{\mathpzc l}_{\Lambda}}^{\zeta_{\cal H}}
+ \langle x \rangle_{{\mathpzc s}_{\Lambda}}^{\zeta_{\cal H}}  \nonumber \\
& :=
\int_0^1 dx\, x [2 {\mathpzc l}_V^{\Lambda}(x;\zeta_{\cal H})
+ {\mathpzc s}_V^{\Lambda}(x;\zeta_{\cal H})] = 1\,. \label{momentumsumrule(Lambda) eq(22b)}
\end{align}
\end{subequations}

\subsection{$\Sigma^0$ baryon -- helicity independent}
Regarding the hadron-scale DF of the $s$ quark in the $\Sigma^0$, one must calculate five terms:
\begin{equation}
{\mathpzc s}_V^{\Sigma}(x;\zeta_{\cal H}) =
\rule{-1em}{0ex}  \sum_{t=Q^{\{ud\}}, D^{[us]},
D^{ [ds]}, D^{\{us\}}, D^{\{ds\}}}
\rule{-1em}{0ex} {\mathpzc s}_{V_t}^{\Sigma}(x;\zeta_{\cal H})\,.
\label{sDF(Sigma) eq(23)}
\end{equation}
Comparing this with Eq.\,\eqref{sDF(Lambda) eq(2)}, one notes two key differences:
the $Q^{[ud]}$ contribution is replaced by $Q^{\{ud\}}$ and, naturally,
the baryon mass and amplitude are now those of the $\Sigma$.

Explicitly, with $\hat\delta_n^{xP} = n\cdot P \delta(n\cdot \ell - x n\cdot P)$, $n^2=0$, $n\cdot P=-m_{\Sigma}$ in the $\Sigma$ rest frame:
\begin{align}
\Lambda_+ & \gamma\cdot n  {\mathpzc s}_{V_{Q^{\{ud\}}}}^{\Sigma}(x;\zeta_{\cal H}) \Lambda_+   =
\int\! \tfrac{d^4\ell}{(2\pi)^4} \,
\hat\delta_n^{xP} \;
\Lambda_+ Q^{\{ud\}}  \Lambda_+\,, \label{sDF1(Sigma) eq(24)}
\end{align}
where
$\Lambda_+ =(m_{\Sigma}-i\gamma \cdot P)/(2m_{\Sigma})$, $P^2=-m_{\Sigma}^2$; and
\begin{align}
Q^{\{ud\}} & =
c^{r_2^1}\, \bar\psi_\rho^{{r_2^1}}(-P) S_{s}(\ell)  \nonumber \\
& \qquad \times  \gamma\cdot n S_{s}(\ell) c^{r_2^1} \psi_\sigma^{{r_2^1}}(P) \Delta_{\rho\sigma}^{{\{ud\}}}(\ell-P)\,,
\label{Q^{ud}_1 eq(25)}
\end{align}
with Eq.\,\eqref{cr21} implicit so that this expression expands to four terms.

Insofar as the remaining contributions in Eq.\,\eqref{sDF(Sigma) eq(23)} are concerned, one need simply use Eqs.\,\eqref{sDF2and3(Lambda) eq(5)}\,--\,\eqref{fDlsav} with the replacements
$r_2^0 \to r_1^1$, $r_3^0 \to r_3^1$, and $m_\Lambda \to m_\Sigma$.

The helicity-independent hadron-scale valence $u$-in-$\Sigma^0$ DF is also obtained as a sum of five terms:
\begin{equation}
{\mathpzc u}_V^{\Sigma}(x;\zeta_{\cal H}) =
\rule{-1em}{0ex} \sum_{t=Q^{[ds]}, Q^{\{ds\}}, D^{\{ud\}}, D^{[us]}, D^{\{us\}}}
\rule{-1em}{0ex}  {\mathpzc u}_{V_t}^{\Sigma}(x;\zeta_{\cal H})\,.
\label{uDF(Sigma)  eq(26)}
\end{equation}
Comparing with Eq.\,\eqref{uDF(Lambda) eq(9)}, the only difference is $D^{[ud]} \to D^{\{ud\}}$; namely, one resolves a $u$ quark in an axialvector diquark instead of a scalar diquark:
\begin{align}
{\mathpzc u}_{V_{D^{\{ud\}}}}^{\Sigma}(x;\zeta_{\cal H}) = \int_x^1\,\frac{dy}{y}\,
{\mathpzc f}_{D^{\{ud\}}}^{\Sigma}(y;\zeta_{\cal H})
{\mathpzc u}_V^{{\{ud\}}}(x/y;\zeta_{\cal H})\,.
\label{uDF3(Sigma)  eq(27)}
\end{align}
Here, ${\mathpzc u}_V^{\{ud\}}(x;\zeta_{\cal H})$ is the valence $u$-quark DF in the $\{ud\}$ diquark, \ref{diquarkDFs}; and the SCI probability density for finding a $\{ud\}$ diquark carrying a light-front fraction $x$ of the $\Sigma$'s momentum is
\begin{subequations}
\begin{align}
\Lambda_+ \gamma \cdot n & {\mathpzc f}_{D^{\{ud\}}}^{\Sigma}(x;\zeta_{\cal H}) \Lambda_+   =
\int\! \tfrac{d^4\ell}{(2\pi)^4} \,
\hat\delta_n^{xP} \;
\bar\Lambda_+ D^{\{ud\}}  \bar\Lambda_+\,,\label{f_{D^{ud}_1}  eq(28a)}\\
D^{\{ud\}} & =   c_{r_2^1}
\bar\psi_\rho^{r_2^1}(P) S_{s}(\ell-P)  \nonumber \\
& \qquad \times
c_{r_2^1} \psi_\sigma^{r_2^1}(-P)
i n\cdot \partial^\ell \Delta_{\rho\sigma}^{{\{ud\}}}(\ell) \,
\label{D^{ud}_1  eq(28b)},
\end{align}
\end{subequations}
where $\bar\Lambda_+ = (m_{\Sigma}+i\gamma \cdot P)/(2m_{\Sigma})$.

The remaining terms are obtained from
Eqs.\,\eqref{T1QuLambda}, \eqref{T2QuLambda}, \eqref{uDF4(Lambda) eq(16)},
\eqref{uDF5(Lambda) eq(17)}
by replacing $r_2^0 \to r_1^1$, $r_3^0\to r_3^1$, and $m_\Lambda \to m_\Sigma$.

The following identity is readily established:
\begin{equation}
{\mathpzc s}_{V_{Q^{\{ud\}}}}^{\Sigma}(x;\zeta_{\cal H}) = {\mathpzc f}_{D^{\{ud\}}}^{\Sigma}(1-x;\zeta_{\cal H}) \,,
\label{sVQfDud}
\end{equation}
as are the $\Sigma^0$ analogues of Eqs.\,\eqref{identity2 eq(20)}, \eqref{identity3 eq(21)}.  Together and combined with the canonical normalisation of the $\Sigma^0$ Faddeev amplitude, these relations guarantee conservation of baryon number and momentum:
\begin{subequations}
\label{SumRules(Sigma)  eq(31)}
\begin{align}
\int_0^1 dx\, & {\mathpzc l}_V^{\Sigma}(x;\zeta_{\cal H})  = 1\,,\;
%
%
\int_0^1 dx\, {\mathpzc s}_V^{\Sigma}(x;\zeta_{\cal H})  = 1\,, \label{baryonnumber(Sigma) eq(31a)}\\
& 2 \langle x \rangle_{{\mathpzc l}_{\Sigma}}^{\zeta_{\cal H}}
+ \langle x \rangle_{{\mathpzc s}_{\Sigma}}^{\zeta_{\cal H}}  \nonumber \\
& :=
\int_0^1 dx\, x [2 {\mathpzc l}_V^{\Sigma}(x;\zeta_{\cal H})
+ {\mathpzc s}_V^{\Sigma}(x;\zeta_{\cal H})] = 1\,. \label{momentumsumrule(Sigma) eq(31b)}
\end{align}
\end{subequations}

\section{Hadron-Scale Helicity-Dependent Parton Distribution Functions -- Algebraic Formulae}
\label{HSHDDFs}
Once again, owing to isospin symmetry, in both $\Lambda, \Sigma^0$, the $u$ and $d$ quark polarised DFs are separately identical; and because quark exchange diagrams are absent from the SCI interaction current, these DFs do not exhibit sensitivity to the relative signs between diquark factors in Eqs.\,\eqref{FaddeevFlavour}.

\subsection{$\Lambda$ baryon -- helicity dependent}
\label{LambdaForm}
The hadron-scale in-$\Lambda$ polarised valence $s$ quark DF receives five contributions:
\begin{equation}
\Delta{\mathpzc s}_V^{\Lambda}(x;\zeta_{\cal H}) =
\rule{-1em}{0ex} \sum_{t=Q^{[ud]}, D^{\{us\}}, D^{\{ds\}}, D^{us}, D^{ds}}
\rule{-1em}{0ex} \Delta{\mathpzc s}_{V_t}^{\Lambda}(x;\zeta_{\cal H})\,.
\label{sDF(Lambda)polarised eq(32)}
\end{equation}
Scalar diquarks cannot be polarised, so here there are no analogues of $D^{[ls]}$ in Eq.\,\eqref{sDF(Lambda) eq(2)}.

{\allowdisplaybreaks
Term~$1_{\Delta{\mathpzc s}}$.  $s$ quark struck with $[ud]$ bystander:
\begin{subequations}
\label{Aafterhere1}
\begin{align}
\Lambda_+ &  \gamma_5\gamma\cdot n  \Delta{\mathpzc s}_{V_{Q^{[ud]}}}^{\Lambda} (x;\zeta_{\cal H}) \Lambda_+  \nonumber \\
& \qquad =  \int\! \tfrac{d^4\ell}{(2\pi)^4} \,
\hat\delta_n^{xP} \;
\Lambda_+ \Delta Q^{[ud]}  \Lambda_+\,,  \label{sDF1(Lambda)polarised eq(33)}\\
\Delta & Q^{[ud]} = [c^{r_1^0}]^2 \,
\bar\psi^{r_1^0}(-P) S_{s}(\ell)  \nonumber \\
& \qquad \times {\mathpzc A}_s \gamma_5 \gamma\cdot n S_{s}(\ell) \psi^{r_1^0}(P) \Delta^{{[ud]}}(\ell-P)\,.
\label{Delta Q^[ud]_0 eq(34)}
\end{align}
\end{subequations}
Here, ${\mathpzc A}_s = 0.695$ is the dressed $s$-quark axial charge.  For quark, $q$, it is obtained as the forward limit of the dressed $q$-quark-axialvector vertex \cite[Eq.\,(A.25)]{Xing:2022sor}:
\begin{equation}
{\mathpzc A}_q = 1/(1 + 4 \overline{\cal C}^{\rm iu}_1(M_q^2)  M_q^2 \alpha_{\rm IR}/[3\pi m_G^2])\,,
\label{Aqcharge}
\end{equation}
with the function $\overline{\cal C}^{\rm iu}_1$ being one of a set that is ubiquitous in SCI expressions:
\begin{align}
n! \, \overline{\cal C}^{\rm iu}_n(\sigma) & = \Gamma(n-1,\sigma \tau_{\textrm{uv}}^{2}) - \Gamma(n-1,\sigma \tau_{\textrm{ir}}^{2})\,,
\label{eq:Cn}
\end{align}
where $\Gamma(\alpha,y)$ is the incomplete gamma-function.  The ``iu'' superscript emphasises that the function depends on both the infrared and ultraviolet cutoffs, $\Lambda_{\rm ir}$, $\Lambda_{\rm uv}$.  Sometimes, the following product is useful:
${\cal C}^{\rm iu}_n(\sigma)=\sigma \overline{\cal C}^{\rm iu}_n(\sigma)$, $n\in {\mathbb Z}^\geq$.
}

Terms $2_{\Delta{\mathpzc s}}$ and $3_{\Delta{\mathpzc s}}$.  $s$-quark within $\{ls\}$ diquarks:
\begin{align}
\Delta{\mathpzc s}_{V_{D^{\{ls\}}}}^{\Lambda}& (x;\zeta_{\cal H})  \nonumber \\
&= \int_x^1\,\frac{dy}{y}\,\Delta{\mathpzc f}_{D^{\{ls\}}}^{\Lambda}(y;\zeta_{\cal H})
{\mathpzc s}_V^{{\{l s\}}}(x/y;\zeta_{\cal H})\,.
\label{sDF2and3(Lambda)polarised eq(35)}
\end{align}
The $\Lambda$--baryon light-front $\{ls\}$ diquark helicity fraction number density is
\begin{subequations}
\label{DeltafD}
\begin{align}
\Lambda_+ & \gamma_5\gamma\cdot n  \Delta{\mathpzc f}_{D^{\{ls\}}}^{\Lambda}(x;\zeta_{\cal H}) \Lambda_+ \nonumber \\
& = \int\! \tfrac{d^4\ell}{(2\pi)^4} \, \hat\delta_n^{xP}\;
\bar \Lambda_+ \Delta D^{\{ls\}}  \bar \Lambda_+\,, \label{Delta f_{D^{qs}_1} eq(36a)} \\
\Delta & D^{\{ls\}}  =   c^{r_3^0} \bar\psi_\rho^{r_3^0}(P)
S_l(\ell-P) \Delta_{\rho\alpha}^{\{ls\}}(\ell) \nonumber \\
&
 \times n_\mu \Gamma_{5\mu;\alpha\beta}^{AAls}(\ell,\ell)\Delta_{\beta\sigma }^{\{ls\}}(\ell)
c^{r_3^0} \psi_\sigma^{r_3^0}(-P)\,, \label{Delta D^{qs}_1 eq(36b)}
\end{align}
\end{subequations}
where Eq.\,\eqref{cr30} is implicit, so there are four terms here, and the axial form factor of the axialvector diquark, $\Gamma_{5\mu;\alpha\beta}^{AAls}$, is given in Ref.\,\cite[Eq.\,(A37b)]{Cheng:2022jxe}.

Terms $4_{\Delta{\mathpzc s}}$ and $5_{\Delta{\mathpzc s}}$.  $s$-quark resolved in the probe-induced $[ls] \leftrightarrow \{ls\}$ transition between diquarks:
\begin{align}
\Delta{\mathpzc s}_{V_{D^{ls}}}^{\Lambda}&(x;\zeta_{\cal H})  \nonumber \\
&= \int_x^1\,\frac{dy}{y}\,\Delta{\mathpzc f}_{01}^{\Lambda}(y;\zeta_{\cal H})
{\mathpzc s}_V^{01}(x/y;\zeta_{\cal H})\,,
\label{sDF4and5(Lambda)polarised eq(37)}
\end{align}
where ${\mathpzc s}_V^{01}(x;\zeta_{\cal H})$ is the valence $s$-quark DF in the scalar-axial\-vector diquark transition, \ref{diquarkDFs}; and
\begin{subequations}
\label{f01transition}
\begin{align}
\Lambda_+ \gamma_5  & \gamma\cdot n  \Delta{\mathpzc f}_{01}^{\Lambda}(x;\zeta_{\cal H}) \Lambda_+
\nonumber \\
& = \int\! \tfrac{d^4\ell}{(2\pi)^4} \, \hat\delta_n^{xP}\;
\bar \Lambda_+ \Delta D_{01}  \bar \Lambda_+\,, \label{Delta f_{0^+1^+} eq(38a)} \\
\Delta D_{01} & =  c^{r_3^0} \bar\psi_\rho^{r_3^0}(P) S_l(\ell-P) \Delta_{\rho\alpha}^{\{ls\}}(\ell)
 \nonumber \\
& \qquad \times
 n_\mu \Gamma_{5\mu;\alpha}^{SAls}(\ell,\ell)
\Delta^{[ls]}(\ell) c^{r_2^0} \psi^{r_2^0}(-P)\, \nonumber \\
&  \quad + c^{r_2^0} \bar\psi^{r_2^0}(P) S_l(\ell-P) \Delta^{[ls]}(\ell) \nonumber \\
&  \qquad \times
n_\mu \Gamma_{5\mu;\alpha}^{SAls}(\ell,\ell) \Delta_{\alpha\rho}^{\{ls\}}(\ell)
c^{r_3^0} \psi_\rho^{r_3^0}(-P) \,, \label{Delta D_{01}  eq(38b)}
\end{align}
\end{subequations}
with Eq.\,\eqref{cr30} implicit, so each piece in the sum expresses two terms, and $\Gamma_{5\mu;\alpha}^{SAls}$ is given in \linebreak Ref.\,\cite[Eq.\,(A39b)]{Cheng:2022jxe}.

The hadron-scale in-$\Lambda$ polarised valence $u$ quark DF receives four contributions:
\begin{equation}
\Delta{\mathpzc u}_V^{\Lambda}(x;\zeta_{\cal H}) = \sum_{t=Q^{ [ds]}, Q^{\{ds\}}, D^{\{us\}}, D^{us}} \Delta{\mathpzc u}_{V_t}^{\Lambda}(x;\zeta_{\cal H})\,.
\label{uDF(Lambda)polarised eq(39)}
\end{equation}

{\allowdisplaybreaks
Term $1_{\Delta {\mathpzc u}}$.  $u$-quark struck with $[ds]$ bystander:
\begin{subequations}
\label{T1uHDL}
\begin{align}
\Lambda_+ \gamma_5 & \gamma\cdot n  \Delta{\mathpzc u}_{V_{Q^{[ds]}}}^{\Lambda} (x;\zeta_{\cal H}) \Lambda_+ \nonumber \\
& =
\int\! \tfrac{d^4\ell}{(2\pi)^4} \,
\hat\delta_n^{xP} \;
\Lambda_+ \Delta Q^{[ds]}  \Lambda_+\,, \label{uDF1(Lambda)polarised eq(40)}\\
\Delta Q^{[ds]} & = [c^{r_2^0}]^2 \,
\bar\psi^{r_2^0}(-P) S_l(\ell)  \nonumber \\
& \qquad \times {\mathpzc A}_l \gamma_5 \gamma\cdot n S_l(\ell) \psi^{r_2^0}(P) \Delta^{[ls]}(\ell-P)\,.
\label{Delta Q^[qs]_0 eq(41)}
\end{align}
\end{subequations}
Here, ${\mathpzc A}_{l=u=d}=0.738$ is the dressed light-quark axial charge, computed using Eq.\,\eqref{Aqcharge} and relevant results from Table~\ref{Tab:DressedQuarks}.
}

Term $2_{\Delta {\mathpzc u}}$.  $u$-quark struck with $\{ds\}$ bystander:
\begin{subequations}
\begin{align}
\Lambda_+ \gamma_5 & \gamma\cdot n  \Delta{\mathpzc u}_{V_{Q^{\{ds\}}}}^{\Lambda} (x;\zeta_{\cal H}) \Lambda_+ \nonumber \\
& =
\int\! \tfrac{d^4\ell}{(2\pi)^4} \,
\hat\delta_n^{xP} \;
\Lambda_+ \Delta Q^{\{ls\}}  \Lambda_+\,, \label{uDF2(Lambda)polarised eq(42)} \\
\Delta Q^{\{ls\}} & = c^{r_3^0} \, \bar\psi_\rho^{r_3^0}(-P) S_l(\ell) {\mathpzc A}_l \gamma_5\gamma\cdot n \nonumber \\
& \qquad \times   S_l(\ell) c^{r_3^0} \psi_\sigma^{r_3^0}(P) \Delta_{\rho\sigma}^{\{ls\}}(\ell-P)\,,
\label{Delta Q^{qs}_1 eq(43)}
\end{align}
\end{subequations}
which expands to four terms owing to Eq.\,\eqref{cr30}.

Term $3_{\Delta {\mathpzc u}}$.  $u$-quark struck within $\{us\}$ diquark:
\begin{align}
\Delta{\mathpzc u}_{V_{D^{\{us\}}}}^{\Lambda}&(x;\zeta_{\cal H}) \nonumber \\
& = \int_x^1\,\frac{dy}{y}\,\Delta{\mathpzc f}_{D^{\{ls\}}}^{\Lambda}(y;\zeta_{\cal H})
{\mathpzc l}_V^{{\{ls\}}}(x/y;\zeta_{\cal H})\,,
\label{uDF3(Lambda)polarised eq(44)}
\end{align}
where, as before, ${\mathpzc l}_V^{{\{ls\}}}(x;\zeta_{\cal H})$
is the valence $l$-quark DF in an $\{ls\}$ diquark and
$\Delta{\mathpzc f}_{D^{\{ls\}}}^{\Lambda}(x;\zeta_{\cal H})$ is given in Eq.\,\eqref{DeltafD}.

Term $4_{\Delta {\mathpzc u}}$.  $u$-quark resolved in the probe-induced $[us] \leftrightarrow \{us\}$ diquark transition:
\begin{align}
\Delta{\mathpzc u}_{V_{D^{us}}}^{\Lambda}& (x;\zeta_{\cal H}) \nonumber \\
& = \int_x^1\,\frac{dy}{y}\,\Delta{\mathpzc f}_{01}^{\Lambda}(y;\zeta_{\cal H})
{\mathpzc l}_V^{01}(x/y;\zeta_{\cal H})\,,
\label{uDF4(Lambda)polarised eq(45)}
\end{align}
where ${\mathpzc l}_V^{01}(x;\zeta_{\cal H})$ is the valence $l$-quark DF in the scalar-axial\-vector diquark transition, \ref{diquarkDFs}, and \linebreak $\Delta{\mathpzc f}_{01}^{\Lambda}(x;\zeta_{\cal H})$ is given in Eq.\,\eqref{f01transition}.

\subsection{$\Sigma^0$ baryon -- helicity dependent}
\label{SSSigma0HI}
The hadron-scale in-$\Sigma^0$ polarised valence $s$ quark DF receives five contributions:
\begin{equation}
\Delta{\mathpzc s}_V^{\Sigma}(x;\zeta_{\cal H}) =
\rule{-1em}{0ex} \sum_{t=Q^{\{ud\}} ,D^{\{us\}}, D^{\{ds\}}, D^{us}, D^{ds}}
\rule{-1em}{0ex} \Delta{\mathpzc s}_{V_t}^{\Sigma}(x;\zeta_{\cal H})\,.
\label{sDF(Sigma)polarised eq(47)}
\end{equation}
Compared with Eq.\,\eqref{sDF(Lambda)polarised eq(32)}, then apart from the change of mass ($m_\Lambda \to m_\Sigma$) and Faddeev amplitude ($r_2^0 \to r_1^1$, $r_{3}^0 \to r_{3}^1$) in Eqs.\,\eqref{sDF2and3(Lambda)polarised eq(35)}\,--\,\eqref{f01transition}, one need only replace the
$s$ quark struck with $[ud]$ bystander by $s$ quark struck with $\{ud\}$:
\begin{subequations}
\begin{align}
\Lambda_+ \gamma_5&\gamma\cdot n  \Delta{\mathpzc s}_{V_{Q^{\{ud\}}}}^{\Sigma} (x;\zeta_{\cal H}) \Lambda_+  \nonumber \\
&=
\int\! \tfrac{d^4\ell}{(2\pi)^4} \,
\hat\delta_n^{xP} \;
\Lambda_+ \Delta Q^{\{ud\}}  \Lambda_+\,, \label{sDF1(Sigma)polarised eq(48)} \\
\Delta Q^{\{ud\}} & = c^{r_2^1} \, \bar\psi_\rho^{r_2^1}(-P) S_{s}(\ell) {\mathpzc A}_s \gamma_5\gamma\cdot n   \nonumber \\
& \qquad \times  S_{s}(\ell) c^{r_2^1} \psi_\sigma^{r_2^1}(P) \Delta_{\rho\sigma}^{\{ud\}}(\ell-P)\,.
\label{Delta Q^{ud}_1 eq(49)}
\end{align}
\end{subequations}

The hadron-scale in-$\Sigma^0$ polarised valence $u$ quark DF also receives five contributions:
\begin{equation}
\Delta{\mathpzc u}_V^{\Sigma}(x;\zeta_{\cal H}) =
\rule{-1em}{0ex} \sum_{t=Q^{ [ds]}, Q^{\{ds\}}, D^{\{ud\}}, D^{\{us\}}, D^{us}}
\rule{-1em}{0ex} \rule{-1em}{0ex} \Delta{\mathpzc u}_{V_t}^{\Sigma}(x;\zeta_{\cal H})\,.
\label{uDF(Sigma)polarised eq(50)}
\end{equation}
Compared with Eq.\,\eqref{uDF(Lambda)polarised eq(39)}, the $D^{\{ud\}}$ contributions is additional.
It resolves the $u$-quark within the $\{ud\}$ diquark:
\begin{align}
\Delta & {\mathpzc u}_{V_{D^{\{ud\}}}}^{\Sigma}(x;\zeta_{\cal H}) \nonumber \\
& = \int_x^1\,\frac{dy}{y}\,\Delta{\mathpzc f}_{D^{\{ud\}}}^{\Sigma}(y;\zeta_{\cal H})
{\mathpzc l}_V^{{\{ud\}}}(x/y;\zeta_{\cal H})\,,
\label{uDF3(Sigma)polarised eq(51)}
\end{align}
where
\begin{subequations}
\begin{align}
 \Lambda_+ \gamma_5& \gamma\cdot n  \Delta{\mathpzc f}_{D^{\{ud\}}}^{\Sigma}(x;\zeta_{\cal H}) \Lambda_+ \nonumber \\
& = \int\! \tfrac{d^4\ell}{(2\pi)^4} \, \hat\delta_n^{xP}\;
\bar \Lambda_+ \Delta D^{\{ud\}}  \bar \Lambda_+\,, \label{Delta f_{D^{ud}} eq(52a)} \\
\Delta D^{\{ud\}} & =   c^{r_2^1}
\bar\psi_\rho^{r_2^1}(P) S_s(\ell-P) \Delta_{\rho\alpha}^{{\{ud\}}}(\ell)
n_\mu \Gamma_{5\mu;\alpha\beta}^{AAud}(\ell,\ell)\nonumber \\
&
 \qquad\times \Delta_{\beta\sigma }^{{\{ud\}}}(\ell)
c^{r_2^1} \psi_\sigma^{r_2^1}(-P)\,, \label{Delta D^{ud}_1 eq(52b)}
\end{align}
\end{subequations}
with $\Gamma_{5\mu;\alpha\beta}^{AAud}$ given in Ref.\,\cite[Eq.\,(A37b)]{Cheng:2022jxe}.

The remaining entries in Eq.\,\eqref{uDF3(Sigma)polarised eq(51)} are obtained from Eqs.\,\eqref{T1uHDL}\,--\,\eqref{uDF4(Lambda)polarised eq(45)} following the replacements $m_\Lambda \to m_\Sigma$, $r_2^0 \to r_1^1$, $r_3^0\to r_3^1$.

\begin{figure}[t]
\leftline{\large\sf A}
\vspace*{-3ex}

\centerline{%
\includegraphics[clip, width=0.95\linewidth]{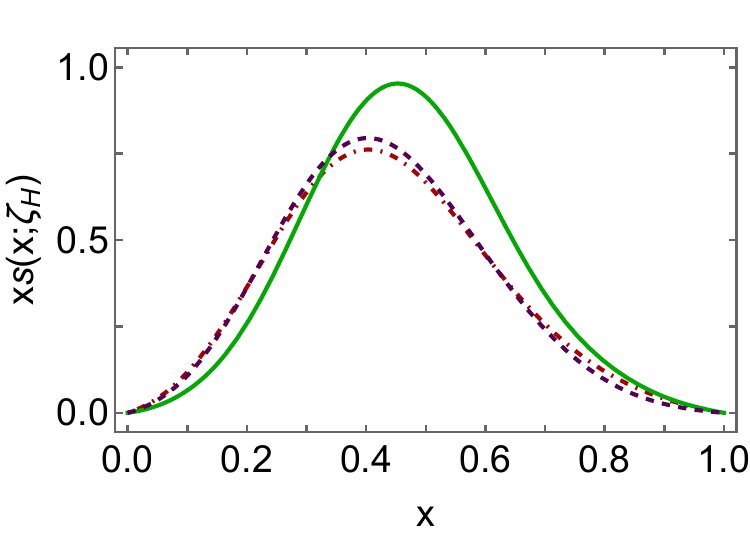}}

\vspace*{1ex}

\leftline{\large\sf B}
\vspace*{-3ex}

\centerline{%
\hspace*{-1ex}\includegraphics[clip, width=0.96\linewidth]{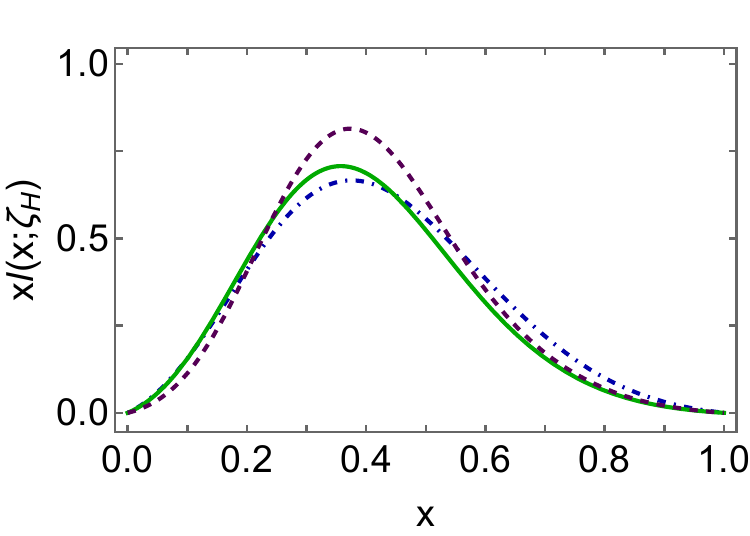}}

\caption{\label{FigUnPolarisedzH}
Hadron scale helicity-independent valence quark DFs.
{\sf Panel A}.
$s$ quark in $\Lambda$ -- solid green curve;
$s$ quark in $\Sigma^0$ -- dashed purple curve;
$0.5\times d$ quark in neutron -- dot-dashed red curve.
{\sf Panel B}.
$l$ quark in $\Lambda$ -- solid green curve;
$l$ quark in $\Sigma^0$ -- dashed purple curve;
$u$ quark in neutron -- dot-dashed blue curve.
(Neutron results from Ref.\,\cite{Yu:2024qsd}.)
}
\end{figure}

\section{Distribution Functions -- Hadron Scale}
\label{DFresultsHS}
\subsection{Numerical Results}
Employing the material in Sect.\,\ref{SolveFE} to complete the formulae collected in \ref{DFappendix}, obtained after evaluating the expressions in Sects.\,\ref{HSHIDFs}, \ref{HSHDDFs}, one arrives at numerical results for all SCI $\Lambda$, $\Sigma^0$ hadron-scale DFs.  Each is accurately interpolated by the following functional form:
\begin{align}
\label{InterpolateDFs}
{\mathpzc f}(x;\zeta_{\cal H}) & = {\mathpzc n}_{\mathpzc f} \, 6 (1-x) \sum_{n=0}^{10}a_n^{\mathpzc f} C_n^{3/2}(1-2x)\,,
\end{align}
where $a_0^{\mathpzc f}=1$ and the remaining coefficients are listed in Tables~\ref{IcoeffsLambda}, \ref{IcoeffsSigma}.

\begin{figure}[t]
\leftline{\large\sf A}
\vspace*{-3ex}

\centerline{%
\includegraphics[clip, width=0.95\linewidth]{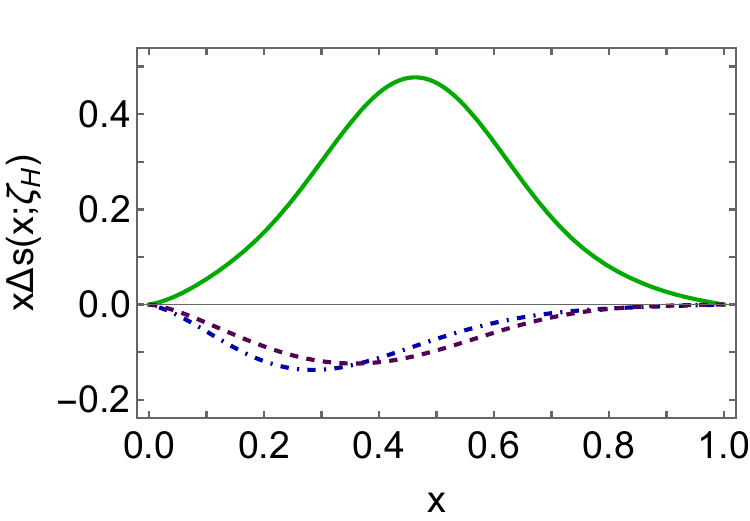}}

\vspace*{1ex}

\leftline{\large\sf B}
\vspace*{-3ex}

\centerline{%
\hspace*{-1ex}\includegraphics[clip, width=0.96\linewidth]{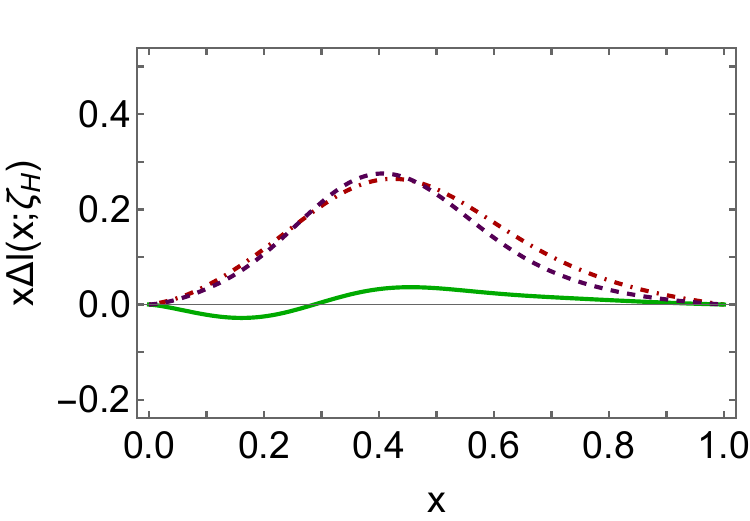}}

\caption{\label{FigPolarisedzH}
Hadron scale helicity-dependent valence quark DFs.
{\sf Panel A}.
$s$ quark in $\Lambda$ -- solid green curve;
$s$ quark in $\Sigma^0$ -- dashed purple curve;
$u$ quark in neutron -- dot-dashed  blue curve.
{\sf Panel B}.
$l$ quark in $\Lambda$ -- solid green curve;
$l$ quark in $\Sigma^0$ -- dashed purple curve;
$0.5\times d$ quark in neutron -- dot-dashed  red curve.
(Neutron results from Ref.\,\cite{Yu:2024qsd}.)
}
\end{figure}

The DFs described above are drawn in Figs.\,\ref{FigUnPolarisedzH}, \ref{FigPolarisedzH}.
Images comparable to Figs.\,\ref{FigUnPolarisedzH}\,A, B are available in Ref.\,\cite{Peng:2022lte}, which solved a light-front model Hamiltonian in its leading Fock sector using a basis light-front quantization (BLFQ) framework.
There are quantitative differences, most notably in the peak locations and heights, which may be attributed to both (\emph{i}) the presence of diquark correlations in our analysis and the absence of such structures in the BLFQ study and (\emph{ii}) the resolving-scale parameter fitting procedure adopted in the BLFQ computations.

\subsection{General observations}

Naturally, the sum rules in Eqs.\,\eqref{SumRules(Lambda) eq(22)}, \eqref{SumRules(Sigma)  eq(31)} are satisfied.
Furthermore, it is explicit in Eq.\,\eqref{InterpolateDFs} and apparent from Figs.\,\ref{FigUnPolarisedzH}, \ref{FigPolarisedzH} that all SCI valence-quark DFs have the same large-$x$ behaviour:
\begin{equation}
\label{SCIpower}
\mbox{SCI:} \quad {\mathpzc f}(x;\zeta_{\cal H}) \stackrel{x\simeq 1}{\propto} (1-x)^1 .
\end{equation}
As in QCD, this power-law is one greater than that on the SCI valence-quark DFs for pseudoscalar mesons \cite{Brodsky:1994kg, Yuan:2003fs, Holt:2010vj, Cui:2021mom, Cui:2022bxn, Lu:2022cjx}:
\begin{align}
\mbox{QCD:} \quad {\mathpzc f}^p(x;\zeta_{\cal H}) & \stackrel{x\simeq 1}{\propto} (1-x)^3 \nonumber \\
& \quad {\emph cf}.\quad {\mathpzc f}^\pi(x;\zeta_{\cal H}) \stackrel{x\simeq 1}{\propto} (1-x)^2\,.
\label{LargeX}
\end{align}
At the other endpoint, as also for the nucleon \cite{Yu:2024qsd}:
\begin{equation}
\mbox{SCI:} \quad {\mathpzc f}(x;\zeta_{\cal H}) \stackrel{x\simeq 0}{\propto} x^0\,, {\rm \emph{i.e}.,~nonzero~constant.}
\end{equation}

That the $\zeta_{\cal H}$ unpolarised and polarised valence quark DFs each separately exhibit the same large-$x$ power law behaviour is consistent with QCD-based expectations.

On the other hand, in large-$\zeta$ DF phenomenology, it is usually argued that there is no correlation between the helicity of the struck quark and that of the parent hadron and that this forces the polarised:unpolarised ratio of DFs to vanish as $x \to 0$.
Using the SCI, this is neither the case for the nucleon nor for $\Lambda$, $\Sigma^0$.
For the proton and neutron at $\zeta_{\cal H}$:
\begin{subequations}
\label{Deltapn}
\begin{align}
\Delta {\mathpzc u}/{\mathpzc u}|_{x\simeq 0}^p & = \phantom{-}0.30\,,\quad
\Delta {\mathpzc d}/{\mathpzc d}|_{x\simeq 0}^p = -0.35 \,, \\
\Delta {\mathpzc u}/{\mathpzc u}|_{x\simeq 0}^n & = -0.35\,, \quad
\Delta {\mathpzc d}/{\mathpzc d}|_{x\simeq 0}^n = \phantom{-}0.30\,.
\end{align}
\end{subequations}
(These values were inadvertently interchanged in Ref.\,\cite{Yu:2024qsd}.)
Turning to the $\Lambda$ and $\Sigma^0$ baryons:
\begin{subequations}
\begin{align}
\Delta {\mathpzc l}/{\mathpzc l}|_{x\simeq 0}^{\Sigma^0} & = \phantom{-} 0.20\,,\quad
\Delta {\mathpzc s}/{\mathpzc s}|_{x\simeq 0}^{\Sigma^0} = -0.29\,, \\
\Delta {\mathpzc l}/{\mathpzc l}|_{x\simeq 0}^\Lambda & = -0.13\,,\quad
\Delta {\mathpzc s}/{\mathpzc s}|_{x\simeq 0}^\Lambda = \phantom{-} 0.82 \,.
\end{align}
\end{subequations}
Moreover, as explained elsewhere -- see Ref.\,\cite[Eq.\,(37)]{Yu:2024qsd} -- this ratio is invariant under evolution; so, it maintains the same nonzero value $\forall \zeta$.  In fact, this is also true for the $x$-dependence of these ratios \cite[Eq.\,(37)]{Yu:2024qsd}, \emph{viz}.\
\begin{equation}
\forall \zeta \geq \zeta_{\cal H}, \;
\Delta {\mathpzc q}(x;\zeta)/{\mathpzc q}(x;\zeta) = \Delta {\mathpzc q}(x;\zeta_{\cal H})/{\mathpzc q}(x;\zeta_{\cal H})\,.
\label{ratioHDHI}
\end{equation}

Regarding Figs.\,\ref{FigUnPolarisedzH}, \ref{FigPolarisedzH}, it is worth highlighting the following collection of similarities at $\zeta_{\cal H}$:
\begin{subequations}
\label{DFsims}
\begin{align}
x{\mathpzc s}^{\Sigma^0}(x) & \quad \sim \quad 0.5 x{\mathpzc d}^n(x)=0.5 x{\mathpzc u}^p(x)\,,
\label{Sim1}\\
x{\mathpzc l}^{\Lambda}(x) & \quad \sim \quad x{\mathpzc u}^n(x)= x{\mathpzc d}^p(x) \,,
\label{Sim2}\\
x\Delta {\mathpzc s}^{\Sigma^0}(x) & \quad \sim \quad x\Delta {\mathpzc u}^{n}(x)=x\Delta {\mathpzc d}^{p}(x) \,, \label{Sim3} \\
x\Delta {\mathpzc l}^{\Sigma^0}(x) & \quad  \sim \quad 0.5x\Delta {\mathpzc d}^{n}(x)=0.5x\Delta {\mathpzc u}^{p}(x) \,. \label{Sim4}
\end{align}
\end{subequations}

%
%
%

Equation \eqref{Sim1} follows because the dominant contribution to the $\Sigma^0$ baryon DF is from the $s$-quark in $[ls]$ and those to the proton DF are $u$ with $[ud]$ and $u$ in $[ud]$.
Scalar diquarks are the lightest, so the mass balance is similar in both cases; hence, in the $x$-weighted DFs, so are the peak locations and domains of material support.
Regarding $s$-in-$\Lambda$, the dominant component is $s[ud]$, \emph{i.e}., the heavier $s$ quark paired with the light diquark.
Consequently, the $s$ carries more of the $\Lambda$-baryon's momentum, shifting the $x\times$\,DF peak to a larger value of $x$.

Looking at $x {\mathpzc l}^{\Lambda}(x)$, Eq.\,\eqref{Sim2}, here the dominant contributions are again those with scalar diquarks: $l$ with $[l^\prime s]$ and $l$ in $[ls]$.
This is quite different from $s$-in-$\Lambda$ and, in any event, given momentum conservation, the peak of this $x\times$\,DF must be shifted to lower $x$ than that of $x{\mathpzc s}^{\Lambda}(x)$.
The DF of the singly-represented quark in the nucleon, say $d$-in-$p$, receives its greatest contribution from the $d$-in-$[ud]$ term, but its profile is modulated by $d$ with $\{ud\}$.
Now, the total mass of $l+[l^\prime s] \approx d+\{ud\}$, so this modulation pushes the peak of the $d$-in-$p$ $x\times\,$DF to lower values of $x$, approaching the vicinity of that in $x{\mathpzc l}^{\Lambda}(x)$.
Hence, the Eq.\,\eqref{Sim2} similarity results from a somewhat complex interplay of complementary mass scales.
%
%
In contrast, the dominant contributions to ${\mathpzc l}^{\Sigma^0}(x)$ are $l$ with $\{l^\prime s\}$ and $l$ in $[ls]$ and the first term here skews the $x\times\,$DF profile to larger values of $x$.

Turning to $\Delta {\mathpzc s}^{\Sigma^0}(x)$, Eq.\,\eqref{Sim3}, the dominant contributions are delivered by the $[ls]\leftrightarrow\{ls\}$ transitions and they are negative.  The same is true for the singly represented quark ($u$) in the neutron.

On the other hand, the dominant contribution to $\Delta {\mathpzc s}^{\Lambda}(x)$ is provided by $s[ud]$.
It is positive because the $[ud]$ cannot be polarised.
Consequently, as has long been anticipated \cite{Boros:1999da}, a large part, but not all, of the $\Lambda$-baryon's helicity is carried by the $s$ quark: not all because of the second scalar diquark component in Eq.\,\eqref{FALambdaA}.

Considering $\Delta {\mathpzc l}^{\Sigma^0}(x)$, Eq.\,\eqref{Sim4}, the dominant contribution is produced by $l[l^\prime s]$.  The scalar diquark cannot be polarised, so the $l$ quark carries the greatest part of the $\Sigma^0$-baryon's helicity.  The same statements are true for the doubly-represented quark in the nucleon.
Turning to $\Delta {\mathpzc l}^{\Lambda}(x)$, there are four terms:
$l[l^\prime s]$ delivers a large positive contribution; $l\{l^\prime s\}$ is negative and small; $l$ in $\{ls\}$ is positive and small; and $[ls]\leftrightarrow\{ls\}$ is negative and large, like in the $\Sigma^0$.  The net effect is almost complete cancellation, which explains the green curve in Fig.\,\ref{FigPolarisedzH}\,B.
However, as we indicate below, this outcome may be model dependent.

Following these remarks, it is evident that the spin-flavour wave functions of the $\Lambda$, $\Sigma^0$ baryons are largely responsible for the character of the curves in Figs.\,\ref{FigUnPolarisedzH}, \ref{FigPolarisedzH}.  Hence, one should anticipate that similar images and conclusions will emerge from QCD-kindred quark + diquark pictures of baryon structure, \emph{viz}.\ studies that use Schwinger functions that more realistically express basic features of QCD \cite{Lu:2022cjx, Cheng:2023kmt}, such as momentum-dependent running quark masses, diquark amplitudes, and baryon Faddeev amplitudes.
There is one caveat, however.
In all studies to date, the photon-baryon DF interaction current has excluded the quark exchange diagram -- see Ref.\,\cite[Fig.\,4]{Cheng:2022jxe} -- and, where relevant, the partner diagrams \cite[Figs.\,5, 6]{Cheng:2022jxe}.
We estimate that the principal impact of these diagrams would be to shift some helicity into light quarks within the $\Lambda$.  More complete analyses should be undertaken in future.

{\allowdisplaybreaks
It is worth recording the hadron-scale valence-quark momentum and helicity fractions in the $\Lambda$, $\Sigma^0$ baryons:
\begin{equation}
\begin{array}{l|ccc}
 & \Lambda & \Sigma^0 & p\  \\\hline
\langle x \rangle_{\mathpzc s}^{\zeta_{\cal H}}\  & \ 0.394\  &  0.345\  & 0\ \\
\langle x \rangle_{\mathpzc u}^{\zeta_{\cal H}}\  & \ 0.303\  & 0.328\  & 2\times 0.343\ \\
\langle x \rangle_{\mathpzc d}^{\zeta_{\cal H}}\  & \ 0.303\  & 0.328\  & \phantom{2\times\ } 0.314\ \\\hline
\langle 1 \rangle_{\Delta{\mathpzc s}}^{\zeta_{\cal H}}/g_A\  & \phantom{-}0.593\ & -0.215\ & 0 \\
\langle 1 \rangle_{\Delta{\mathpzc u}}^{\zeta_{\cal H}}/g_A\  & -0.018\ & \phantom{-}0.347\  & 2\times 0.374\ \\
\langle 1 \rangle_{\Delta{\mathpzc d}}^{\zeta_{\cal H}}/g_A\  & -0.018\ & \phantom{-}0.347\ & \!\! \phantom{2}-0.253 \\
a_0/g_A\  & \!\! \phantom{-}0.56\  & \!\!\phantom{-}0.48\ & \phantom{2\times}0.49\ \\
a_3/g_A\  & 0\ & 0\ & 1\ \\
a_8/g_A\  & -1.22\ & \!\!\phantom{-}1.12\ & \phantom{2\times}0.49\ \\ \hline
\end{array}
\label{TableMoments}
\end{equation}
where the axial charges are ($B=\Lambda, \Sigma^0, p$):
\begin{subequations}
\begin{align}
a_0^B & =
   \langle 1 \rangle_{\Delta{\mathpzc u}}^{\zeta_{\cal H},B} + \langle 1 \rangle_{\Delta{\mathpzc d}}^{\zeta_{\cal H},B} + \langle 1 \rangle_{\Delta{\mathpzc s}}^{\zeta_{\cal H},B} \,, \\
  a_3^B & =
   \langle 1 \rangle_{\Delta{\mathpzc u}}^{\zeta_{\cal H},B} - \langle 1 \rangle_{\Delta{\mathpzc d}}^{\zeta_{\cal H},B} \,,  \\
  a_8^B & =
   \langle 1 \rangle_{\Delta{\mathpzc u}}^{\zeta_{\cal H},p} + \langle 1 \rangle_{\Delta{\mathpzc d}}^{\zeta_{\cal H},p} -2  \langle 1 \rangle_{\Delta{\mathpzc s}}^{\zeta_{\cal H},B}\,,
\end{align}
\end{subequations}
and $a_3^p = g_A$, \emph{i.e}., the nucleon axial charge.  The SCI value is $g_A = 0.92$.
In common with Ref.\,\cite{Yu:2024qsd} as remarked above, herein, a static approximation is employed in treating baryon Faddeev equations and axialvector currents.  Hence, contributions associated with Ref.\,\cite[Fig.\,2-Diagram~4]{Cheng:2022jxe} are omitted; so, $g_A$ is underestimated.
Notwithstanding these remarks, it is worth noting that the SCI predictions for $a_0^{\Sigma^0}/a_0^\Lambda = 0.86$, $a_0^{p}/a_0^\Lambda = 0.89$, $a_0^{\Sigma^0}/a_0^p = 0.97$ are consistent with the lQCD estimates in Ref.\,\cite{CSSM:2014uyt}.
($a_0$ is scale invariant under AO evolution \cite{Yin:2023dbw}.)
}

\subsection{Far valence domain}
On $x\simeq 1$, ratios of DFs are invariant under QCD evolution \cite{Holt:2010vj}.  Consequently, they are definitive, scale invariant characteristics of any analysis and framework.  It is therefore worth recording an array of such ratios as a reference for future comparisons.   (Recall, $l$ is the light valence quark in the $\Lambda$ or $\Sigma^0$ baryon and $s$ is the strange valence quark.)
\begin{equation}
\label{FarVx}
\begin{array}{l|ccl}
 x\simeq 1\  & \ \Lambda\  & \Sigma^0\ & \rule{2em}{0ex} p\  \\\hline
 l/s  & \ 0.36\ & \phantom{-}0.90\ & \\
 d/u &  & & \phantom{-}0.71/2= \phantom{-}0.36 \  \\
 \Delta l/\Delta s & \ 0.16\ & -4.43\ & \\
 \Delta d/\Delta u & & & -0.21/2 = -0.11 \ \\
 \Delta s/s & \ 0.64\ & -0.13\ & \\
 \Delta l/l & \ 0.28\ & \phantom{-}0.62\ & \\
\Delta u/u &  &  & \phantom{-}0.52\ \\
\Delta d/d &  &  & -0.16\ \\ \hline
\end{array}
\end{equation}

Reviewing the results in Eq.\,\eqref{FarVx}, it is worth recalling that nonzero in-proton values of $d/u$, $\Delta d/\Delta u$ are only possible because of the presence of axialvector diquarks in the nucleon; see Eq.\,\eqref{protonSF}.  Without the axialvector diquark, the $d$ quark would always be sequestered within a (soft) diquark correlation; hence, not accessible to truly hard probes.

Similarly, in the absence of axialvector diquarks, the $s$ quark in $\Sigma^0$ would be locked within a scalar diquark -- see Eq.\,\eqref{FaddeevFlavourS} -- so invisible to a truly hard probe, in consequence of which, on $x \simeq 1$, $l/s^{\Sigma^0} = \infty = \Delta l/\Delta s^{\Sigma^0}$.
The dominance of scalar diquarks in the $\Sigma^0$ Faddeev amplitude explains the large values of these ratios in Eq.\,\eqref{FarVx}.
Stated otherwise, only with the existence of axialvector diquark correlations can the $s$ quark participate in hard interactions as a valence degree of freedom in the $\Sigma^0$.  Furthermore, in the absence of axialvector diquarks, the $s$ quark could carry none of the $\Sigma^0$ spin. 

It is also worth stressing that the values in the last four rows of Eq.\,\eqref{FarVx} differ from unity.  Such outcomes are inconsistent with the notion of helicity retention in hard scattering processes \cite{Farrar:1975yb, Brodsky:1994kg}.
In addressing this issue now, only nucleon data is available.
For the proton, helicity retention would be impossible unless \linebreak $\Delta {\mathpzc d}^p(x;\zeta_{\cal H})$ were to possess a zero.
Existing data indicate that if such a zero exists, then it lies on $x\gtrsim 0.6$ \cite[HERMES]{HERMES:2004zsh}, \cite[COMPASS]{COMPASS:2010hwr},
\cite[CLAS EG1]{CLAS:2006ozz, CLAS:2008xos, CLAS:2015otq, CLAS:2017qga},
\cite[E06-014]{JeffersonLabHallA:2014mam},
\cite[E99-117]{JeffersonLabHallA:2003joy, JeffersonLabHallA:2004tea}.
To make progress on this issue, data relating to polarised valence quark DFs on $x \gtrsim 0.6$ are desirable.  They exist for the nucleon \cite[CLAS RGC]{E1206109}, \cite[E12-06-110]{Zheng:2006}.
Completed analyses can reasonably be expected within a few years.

\section{Distribution Functions -- Evolved}
\label{EvolvedDFs}
Any experiments that might deliver data which is relevant to extraction of $\Lambda$ or $\Sigma^0$ parton DFs would involve resolving scales $\zeta > \zeta_{\cal H}$.
A typical reference scale is $\zeta=\zeta_2 := 2\,$GeV.
Here, therefore, we describe evolved results obtained using the AO scheme explained in \linebreak Ref.\,\cite{Yin:2023dbw}, which has proved efficacious in many applications, providing, \emph{e.g}.,
unified predictions for all pion, kaon, and proton DFs \cite{Cui:2020tdf, Chang:2022jri, Lu:2022cjx, Cheng:2023kmt, Yu:2024qsd};
insights from experiment into such DFs \cite{Xu:2023bwv, Xu:2024nzp};
useful information on quark and gluon angular momentum contributions to the proton spin \cite{Yu:2024ovn};
predictions for $\pi$ and $K$ fragmentation functions \cite{Xing:2025eip};
and a tenable species separation of nucleon gravitational form factors \cite{Yao:2024ixu}.

All-orders evolution is based on two axioms.
The first is the concept of an effective charge \cite{Grunberg:1980ja, Grunberg:1982fw, Deur:2023dzc}; namely, a QCD running coupling defined by any single observable via the formula which expresses that observable to first-order in the perturbative coupling.
By definition, such a coupling incorporates terms of arbitrarily high order in the perturbative coupling and (\emph{i}) is consistent with the QCD renormalisation group; (\emph{ii}) renormalisation scheme independent; (\emph{iii}) everywhere analytic and finite; and (\emph{iv}) supplies an infrared completion of any standard running coupling.

Concerning DFs, the AO approach posits that there is at least one charge, $\alpha_{1\ell}(k^2)$, which, when used to integrate the leading-order perturbative DGLAP equations \cite{Dokshitzer:1977sg, Gribov:1971zn, Lipatov:1974qm, Altarelli:1977zs}, defines an evolution scheme for \emph{every} DF that is all-orders exact.  This definition is unusually broad because it refers to an entire class of observables.
It is worth emphasising that the pointwise form of $\alpha_{1\ell}(k^2)$ is largely irrelevant.
Notwithstanding that, the process-independent strong running coupling defined and computed in Refs.\ \cite{Binosi:2016nme, Cui:2019dwv} has all the required properties.

The second plank supporting AO evolution is a definition of the hadron scale, $\zeta_{\cal H}<{\rm mass}_{\rm nucleon}=:m_N$, \emph{i.e}., the starting scale for evolution.
All ambiguity is eliminated when one associates $\zeta_{\cal H}$ with that scale at which all properties of a given hadron are carried by its valence degrees of freedom.
This means, for instance, that all of a hadron's light-front momentum is carried by valence degrees of freedom at  $\zeta_{\cal H}$.
Consequently, the DFs associated with glue and sea quarks are identically zero at $\zeta_{\cal H}$.
Working with the running coupling discussed in Refs.\,\cite{Binosi:2016nme, Cui:2019dwv, Deur:2023dzc, Brodsky:2024zev}, the value of the hadron scale is a prediction \cite{Cui:2021mom}:
\begin{equation}
  \zeta_{\cal H} = 0.331(2)\,{\rm GeV}\,.
\end{equation}
This value is confirmed in an analysis of lQCD studies of the pion valence quark DF \cite{Lu:2023yna}.

Regarding the evolution of singlet DFs in the proton, a Pauli blocking factor is included in the gluon splitting function \cite[Sec.\,6]{Chang:2022jri}:
\begin{equation}
\label{gluonsplit}
P_{f \leftarrow g}(x;\zeta) \to P_{f \leftarrow g}(x) +
 \sqrt{3}  (1 - 2 x) \frac{ {\mathpzc g}_{f} }{1+(\zeta/\zeta_H-1)^2}\,,
\end{equation}
where $P_{f \leftarrow g}(x) $ is the usual one-loop gluon splitting function and the strength factors are  ${\mathpzc g}_{s,\bar s}=0={\mathpzc g}_{c,\bar c}$, and ${\mathpzc g}_{d,\bar d}= 0.34 =  -{\mathpzc g}_{u,\bar u}=:{\mathpzc g}$.
This modification shifts momentum into $d+\bar d$ from $u+\bar u$, otherwise leaving the total sea momentum fraction unchanged.
Reflecting the waning effect of valence quarks as the proton's glue and sea content increases, it vanishes with increasing $\zeta$.
The modification does not affect baryon number conservation.

\begin{figure}[t]
\leftline{\large\sf A}
\vspace*{-3ex}

\centerline{%
\includegraphics[clip, width=0.95\linewidth]{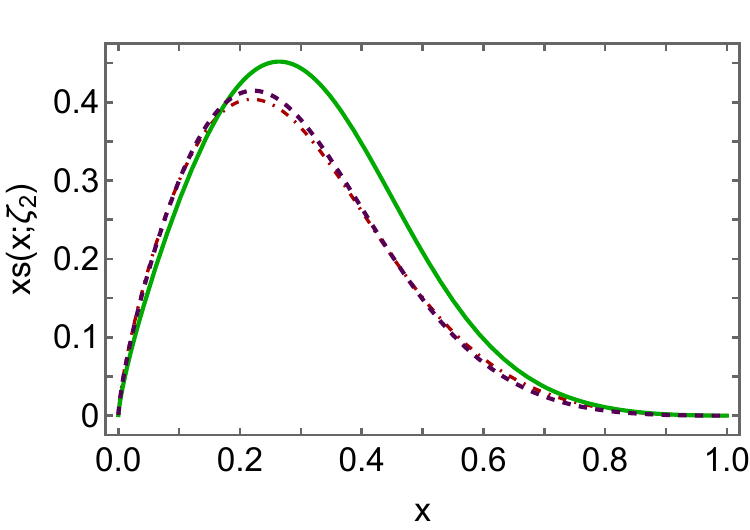}}

\vspace*{1ex}

\leftline{\large\sf B}
\vspace*{-3ex}

\centerline{%
\hspace*{-1ex}\includegraphics[clip, width=0.95\linewidth]{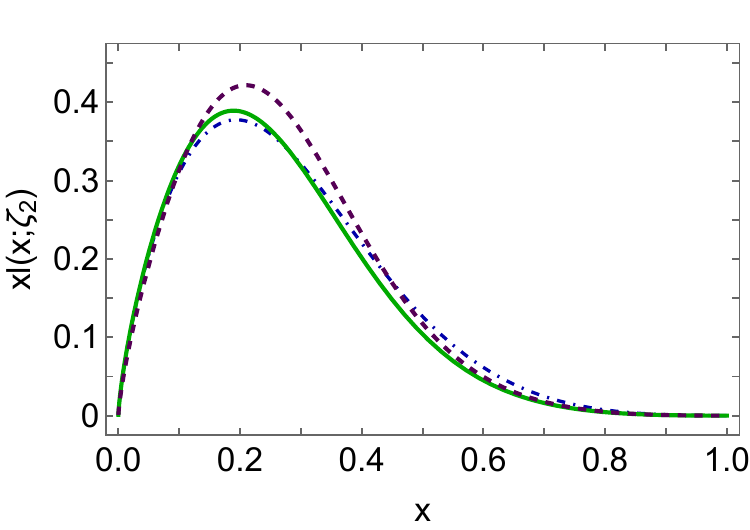}}

\caption{\label{FigUnPolarisedz2}
Helicity-independent valence quark DFs at $\zeta=\zeta_2 := 2\,$GeV.
{\sf Panel A}.
$s$ quark in $\Lambda$ -- solid green curve;
$s$ quark in $\Sigma^0$ -- dashed purple curve;
$0.5\times d$ quark in neutron -- dot-dashed red curve.
{\sf Panel B}.
$l$ quark in $\Lambda$ -- solid green curve;
$l$ quark in $\Sigma^0$ -- dashed purple curve;
$u$ quark in neutron -- dot-dashed blue curve.
(Neutron results from Ref.\,\cite{Yu:2024qsd}.)
}
\end{figure}

Since $\Lambda$, $\Sigma^0$ baryons do not contain a doubly represented flavour, no such factor is included in the evolution equations for their DFs.
This would change, \emph{e.g}., when considering $\Sigma^\pm$, $\Xi^{-,0}$ baryons.

In all cases we implement a flavour threshold effect, multiplying the glue$\,\to\,$quark splitting function by the factor \cite{Lu:2022cjx}:
\begin{equation}
\label{MassDependent}
 {\cal P}_{qg}^\zeta = \tfrac{1}{2} \left(1+\tanh[ (\zeta^2 - \delta_q^2)/\zeta_{\cal H}^2] \right)\,,
\end{equation}
$\delta_{u,d}\approx 0$, $\delta_s \approx 0.1\,$GeV, $\delta_c \approx 0.9\,$GeV.
This threshold function guarantees that a given quark flavour only participates in DF evolution after the resolving energy scale exceeds a value determined by the quark's mass.

\subsection{Helicity independent: $\zeta = \zeta_2$ predictions}
\label{HIz2predictions}
Figure~\ref{FigUnPolarisedz2} displays the hadron-scale valence-quark curves in Fig.\,\ref{FigUnPolarisedzH} after evolution  to $\zeta=\zeta_2$.
Evolution relocates support in valence quark DFs to lower $x$; hence, whilst it preserves the patterns of difference and similarity described previously, they steadily become less perceptible with increasing scale as the DF support domain is compressed.
Given known uncertainties in DF inferences from data, Fig.\,\ref{FigUnPolarisedz2} highlights that very precise data and reliable phenomenology would be necessary before the predicted differences between $\Lambda$, $\Sigma^0$, $p$ DFs could be confirmed empirically.

An effective large-$x$ exponent for the $\zeta_2$ SCI DFs in Fig.\,\ref{FigUnPolarisedz2} can readily be obtained.
Namely, focusing on the domain $0.85< x < 1$, one performs a least-squares best-fit in the form ${\mathpzc c} (1-x)^\beta$ and thereby obtains
\begin{equation}
{\rm HI:} \quad \beta_{\rm valence}^{\rm SCI} \stackrel{0.85 < x < 1}{=} 2.35(5)\,.
\label{largexHIvalence}
\end{equation}
Analyses using Schwinger functions with QCD-like momentum dependence yield \cite{Lu:2022cjx} $ \beta_{\rm valence} \approx 4$.

\begin{figure}[t]
\centerline{%
\includegraphics[clip, width=0.95\linewidth]{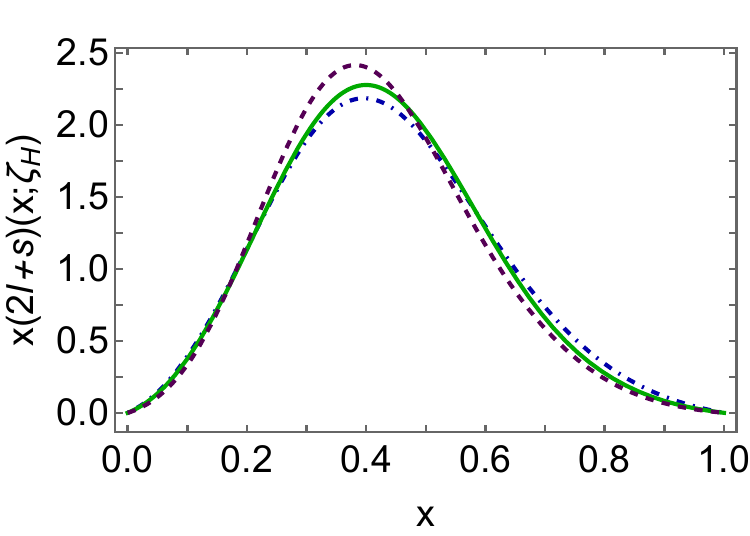}}

\caption{\label{GlueSeaSources}
Sum of $\zeta=\zeta_{\cal H}$ valence-quark DFs in $\Lambda$ -- solid green curve; $\Sigma^0$ -- dashed purple curve; and $p$ -- dot-dashed blue curve, which is actually $u+d$.}
\end{figure}

Nonzero glue and sea DFs emerge on $\zeta>\zeta_{\cal H}$ following AO evolution.  They are generated from the sum of $\zeta=\zeta_{\cal H}$ valence-quark DFs in the hadron.  Regarding $\Lambda$, $\Sigma^0$, $p$, these source profiles are displayed in Fig.\,\ref{GlueSeaSources}.  Evidently, there are discernible differences between the $x$-weighted DFs in a neighbourhood of $x=0.4$, but even they are not large.  One should therefore expect glue and sea DFs in these hadrons to be similar.

\begin{figure}[t]
\leftline{\large\sf A}
\vspace*{-3ex}

\centerline{%
\includegraphics[clip, width=0.95\linewidth]{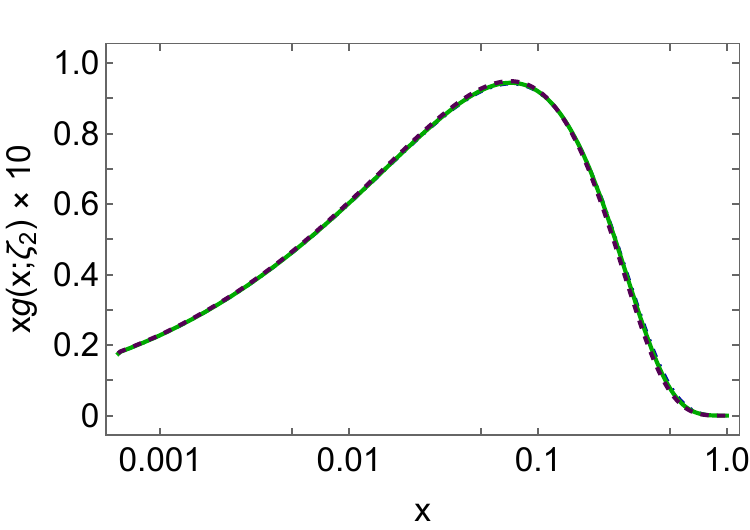}}

\vspace*{1ex}

\leftline{\large\sf B}
\vspace*{-3ex}

\centerline{%
\hspace*{-1ex}\includegraphics[clip, width=0.95\linewidth]{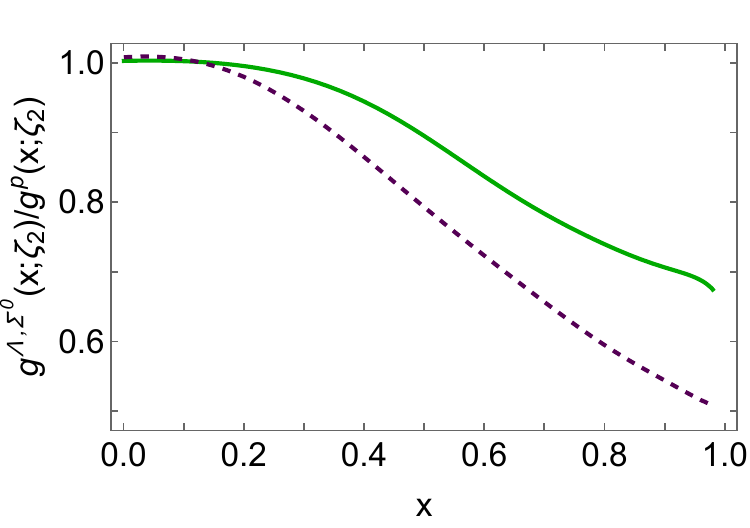}}
\caption{\label{glueDFs}
Helicity-independent glue DFs at $\zeta=\zeta_2$.
{\sf Panel A}.
$g$ in $\Lambda$ -- solid green curve;
in $\Sigma^0$ -- dashed purple curve;
in nucleon -- dot-dashed blue curve.
{\sf Panel B}.
${g}^\Lambda/{g}^p$ -- solid green curve;
and ${g}^{\Sigma^0}/{g}^p$ -- dashed purple curve.}
\end{figure}

Glue DFs are drawn in Fig.\,\ref{glueDFs}.
Regarding Fig.\,\ref{glueDFs}\,A, on the scale of this image, the glue DFs in $\Lambda$, $\Sigma^0$, $p$ are indistinguishable.
Turning to Fig.\,\ref{glueDFs}\,B, which depicts ratios of $\Lambda, \Sigma^0$-to-$p$, one sees that the additional support in the proton valence-quark DFs on $x\gtrsim 0.6$, evident from close scrutiny of Fig.\,\ref{GlueSeaSources}, translates into additional strength in the proton glue DF on the valence-quark domain.
Plainly, since the $x\simeq 1$ values of the ratios are finite, these DFs all have the same large-$x$ power law behaviour.
The effective exponent is
\begin{equation}
\label{largexHIglue}
{\rm HI:} \quad \beta_{\rm glue}^{\rm SCI} \stackrel{0.85 < x < 1}{=} 3.56(3)\,,
\end{equation}
which is roughly $1+\beta_{\rm valence}^{\rm SCI}$, as in QCD.

\begin{figure}[t]
\leftline{\large\sf A}
\vspace*{-3ex}

\centerline{%
\includegraphics[clip, width=0.95\linewidth]{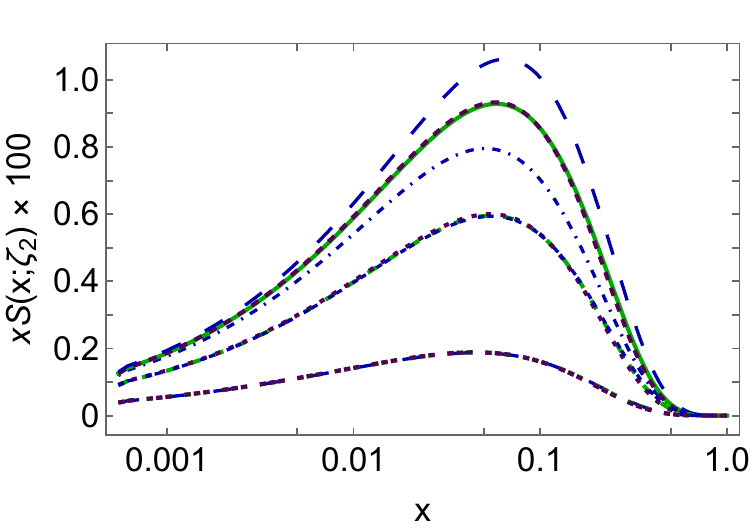}}

\vspace*{1ex}

\leftline{\large\sf B}
\vspace*{-3ex}

\centerline{%
\hspace*{-1ex}\includegraphics[clip, width=0.95\linewidth]{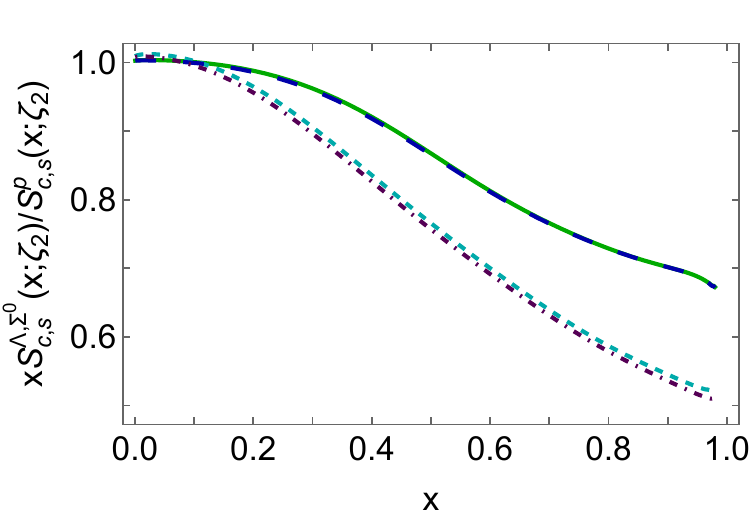}}
\caption{\label{SeaDFs}
Helicity-independent sea DFs at $\zeta=\zeta_2$.
{\sf Panel A}.
Proton (listed in order of $x\simeq 0.1$ magnitudes).
$2\bar d$  -- long-dashed blue curve;
$2\bar u$  -- dot-dashed blue curve;
$s+\bar s$ -- double-dashed blue curve;
$c+\bar c$ -- short-dashed blue curve.
$\Lambda$ (listed in order of $x\simeq 0.1$ magnitudes).
$2\bar l$  -- solid green curve;
$2\bar s$ -- medium-dashed green curve;
$c+\bar c$ -- triple-dashed green curve.
$\Sigma^0$ (listed in order of $x\simeq 0.1$ magnitudes).
$2\bar l$ -- dashed purple curve;
$2\bar s$  -- short-dashed purple curve;
$c+\bar c$ -- triple-dashed purple curve.
{\sf Panel B}.
${\mathpzc S}_{s}^\Lambda/{\mathpzc S}_s^p$ -- solid green curve;
${\mathpzc S}_{c}^\Lambda/{\mathpzc S}_c^p$ -- long-dashed blue curve;
${\mathpzc S}_{s}^{\Sigma^0}/{\mathpzc S}_s^p$ -- short-dashed cyan curve;
${\mathpzc S}_{c}^{\Sigma^0}/{\mathpzc S}_c^p$ -- dot-dashed purple curve.}
\end{figure}

Sea DFs are drawn in Fig.\,\ref{SeaDFs}: qualitatively and semiquantitatively, they are similar to the glue DFs.
The in-proton $\bar d$ \emph{cf}.\ $\bar u$ difference, evident in Fig.\,\ref{SeaDFs}\,A, is driven by Pauli blocking; see Eq.\,\eqref{gluonsplit}.
Regarding Fig.\,\ref{SeaDFs}\,B, the ratio plot for the in-$\Lambda$ light quark sea DFs as compared to the average of in-proton $\bar u, \bar d$ are practically indistinguishable from those drawn; likewise for the analogous in-$\Sigma^0$ ratio.
The effective large-$x$ exponent for SCI sea DFs is
\begin{equation}
{\rm HI:} \quad \beta_{\rm sea}^{\rm SCI} \stackrel{0.85 < x < 1}{=} 4.51(2)\,,
\label{EffectivePowerSea}
\end{equation}
which, in keeping with the QCD DGLAP pattern, is roughly $2+\beta_{\rm valence}^{\rm SCI}$.
It is further worth highlighting that the SCI analysis predicts a nonzero $c+\bar c$ DF in each hadron discussed herein, \emph{viz}.\ $\Lambda, \Sigma^0, p$.
All such $c+\bar c$ DFs have sea quark profiles and fairly commensurate magnitudes.

\begin{table*}[t]
\caption{
\label{MellinMoments}
Low-order $\zeta=\zeta_2$ Mellin moments for the $p, \Lambda, \Sigma^0$ DFs discussed in Sect.\,\ref{HIz2predictions}.  All results reported as percentages.
}
\begin{tabular*}
{\hsize}
{
l@{\extracolsep{0ptplus1fil}}
c@{\extracolsep{0ptplus1fil}}
c@{\extracolsep{0ptplus1fil}}
c@{\extracolsep{0ptplus1fil}}
c@{\extracolsep{0ptplus1fil}}
c@{\extracolsep{0ptplus1fil}}
c@{\extracolsep{0ptplus1fil}}
c@{\extracolsep{0ptplus1fil}}}\hline
%
%
Proton & ${\mathpzc u}^p \ $ & ${\mathpzc d}^p \ $ & ${\mathpzc g}^p\ $ &
${\mathpzc S}_p^u\ $ & ${\mathpzc S}_p^{d}\ $ & ${\mathpzc S}_p^s\ $ & ${\mathpzc S}_p^c\ $ \\ \hline
$\langle x \rangle^{\zeta_2}\ $ & $32.7\phantom{0}\ $ & $15.0\phantom{0}\ $ & $41.4\phantom{10}\ $& $3.29\phantom{10}\ $ & $4.19\phantom{10}\ $ & $2.48\phantom{10}\ $& $0.850\phantom{0}\ $ \\
$\langle x^2\rangle^{\zeta_2}\ $ & $\phantom{1}9.47\ $ & $\phantom{1}4.15\ $ & $2.67\phantom{0}\ $& $0.175\phantom{0}\ $& $0.258\phantom{0}\ $ & $0.136\phantom{0}\ $ & $0.0404\ $\\
$\langle x^3\rangle^{\zeta_2}\ $ & $\phantom{1}3.58\ $ & $\phantom{1}1.54\ $ & $0.500\ $& $0.0267\ $ & $0.0425\ $ & $0.0213\ $ & $0.0060\ $ \\
\hline
$\Lambda$
& ${\mathpzc (u+d)}^{\Lambda} \ $ & ${\mathpzc s}^{\Lambda} \ $ & ${\mathpzc g}^{\Lambda}\ $ &
${\mathpzc S}_{\Lambda}^u\ $ & ${\mathpzc S}_{\Lambda}^{d}\ $ & ${\mathpzc S}_{\Lambda}^s\ $ & ${\mathpzc S}_{\Lambda}^c\ $ \\ \hline
$\langle x \rangle^{\zeta_2}\ $ & $28.9\phantom{0}\ $ & $18.8\phantom{0}\ $ & $41.4\phantom{00}\ $& $3.74\phantom{00}\ $ & $3.74\phantom{00}\ $ & $2.48\phantom{11}\ $& $0.850\ $ \\
$\langle x^2\rangle^{\zeta_2}\ $ & $\phantom{1}7.60\ $ & $\phantom{1}5.83\ $ & $\phantom{1}2.63\phantom{0}\ $& $0.214\phantom{0}\ $& $0.214\phantom{0}\ $ & $\phantom{1}0.134\phantom{11}\ $ & $\phantom{1}0.0399\ $\\
$\langle x^3\rangle^{\zeta_2}\ $ & $\phantom{1}2.66\ $ & $\phantom{1}2.29\ $ & $\phantom{1}0.483\ $& $0.0335\ $ & $0.0335\ $ & $\phantom{11}0.0206\phantom{11}\ $ & $\phantom{1}0.0058\ $ \\\hline
$\Sigma^0$
& ${\mathpzc (u+d)}^{\Sigma^0} \ $ & ${\mathpzc s}^{\Sigma^0} \ $ & ${\mathpzc g}^{\Sigma^0}\ $ &
${\mathpzc S}_{\Lambda}^u\ $ & ${\mathpzc S}_{\Lambda}^{d}\ $ & ${\mathpzc S}_{\Lambda}^s\ $ & ${\mathpzc S}_{\Lambda}^c\ $ \\ \hline
$\langle x \rangle^{\zeta_2}\ $ & $31.3\phantom{0}\ $ & $16.4\phantom{0}\ $ & $41.4\phantom{00}\ $& $3.74\phantom{00}\ $ & $3.74\phantom{00}\ $ & $2.48\phantom{11}\ $& $0.850\ $ \\
$\langle x^2\rangle^{\zeta_2}\ $ & $\phantom{1}8.46\ $ & $\phantom{1}4.70\ $ & $\phantom{1}2.58\phantom{0}\ $& $0.209\phantom{0}\ $& $0.209\phantom{0}\ $ & $\phantom{1}0.132\phantom{11}\ $ & $\phantom{1}0.0391\ $\\
$\langle x^3\rangle^{\zeta_2}\ $ & $\phantom{1}2.99\ $ & $\phantom{1}1.74\ $ & $\phantom{1}0.461\ $& $0.0320\ $ & $0.0320\ $ & $\phantom{11}0.0198\phantom{11}\ $ & $\phantom{1}0.0056\ $ \\
\hline
\end{tabular*}
\end{table*}

Working with the DFs drawn in Figs.\,\ref{FigUnPolarisedz2}\,--\,\ref{SeaDFs}, one obtains the low-order Mellin moments listed in Table~\ref{MellinMoments}.
Unlike many studies that have followed Ref.\,\cite{Jaffe:1980ti}, we make no reference to any other source of moments nor fit any parameters to secure agreement therewith.
Instead, the values are predictions, derived entirely from the hadron-scale DFs in Fig.\,\ref{FigUnPolarisedzH} by using AO evolution \cite{Yin:2023dbw}.
Thus, the results in Table~\ref{MellinMoments} can serve as a ready source of material with which other analyses can compare in future.
In this connection, we suggest that the quenched lQCD study in Ref.\,\cite{Gockeler:2002uh}, which reports the following in-$\Lambda$ momentum fractions:
$\langle x \rangle_{\mathpzc (u+d)}^{\Lambda} =0.40(1)$
$\langle x \rangle_{s}^{\Lambda} =0.27(1)$, should be revisited.
That being the case, then some of the results in Ref.\,\cite{Peng:2022lte}, which were obtained after fitting a parameter to reproduce the quen\-ched lQCD moments, should also be reconsidered.

A general feature of results produced by AO evolution is made apparent by Table~\ref{MellinMoments}; namely, in each hadron constituted from some collection of the lighter three quarks, so long as each flavour radiates glue with equal probability, then the momentum fractions lodged with valence, glue, and total sea are separately identical.  This is simultaneously true for both mesons and baryons; compare, \emph{e.g}., Refs.\,\cite{Cui:2020tdf, Lu:2022cjx} with Table~\ref{MellinMoments}.

It is worth highlighting the impact of Higgs boson couplings into QCD on the predictions in Table~\ref{MellinMoments}.
Note, therefore, that the presence of diquark correlations in the proton Faddeev wave function entails that the light-front fraction of the proton's momentum carried by each $u$ valence quark is 9.2\% more than that carried by the $d$ valence quark.  There is no Higgs boson contribution to this shift.

Consider now the $\Sigma^0$ baryon, in which the diquark structure is similar to that of the nucleon.
The $s$ quark in the $\Sigma^0$ carries $5.2$\% more momentum than each of the light quarks.
This modest effect can largely be attributed to the heavier dressed-quark mass obtained by solving the gap equation using the $s$-quark current mass; see Table~\ref{Tab:DressedQuarks}.
The $s$ current mass is $24$-times larger than the $l$ current mass, but EHM overwhelms this difference, so that the Higgs impact on such observables is strongly suppressed.

Regarding the $\Lambda$ baryon, Higgs boson effects are somewhat amplified by the dominance of lighter scalar diquark correlations in the wave function.  In this case, the $s$ quark carries $30$\% more momentum than each of the light quarks.

\subsection{Helicity dependent: $\zeta = \zeta_2$ predictions}
\label{HDz2predictions}
\begin{figure}[t]
\leftline{\large\sf A}
\vspace*{-3ex}

\centerline{%
\includegraphics[clip, width=0.95\linewidth]{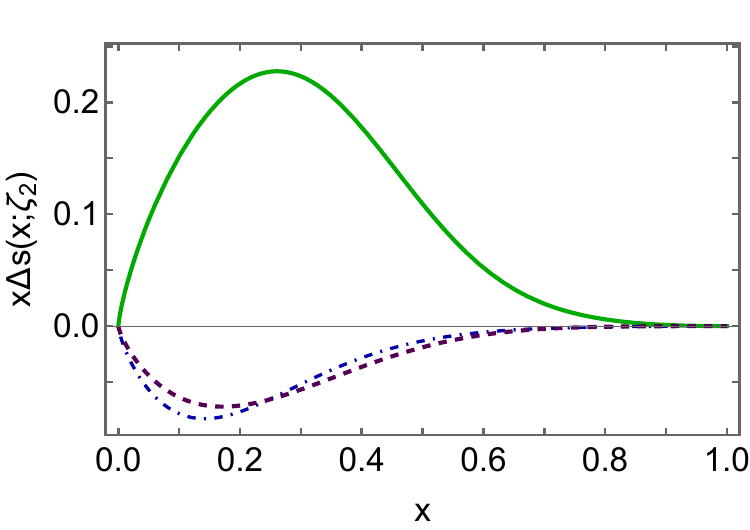}}

\vspace*{1ex}

\leftline{\large\sf B}
\vspace*{-3ex}

\centerline{%
\hspace*{-1ex}\includegraphics[clip, width=0.96\linewidth]{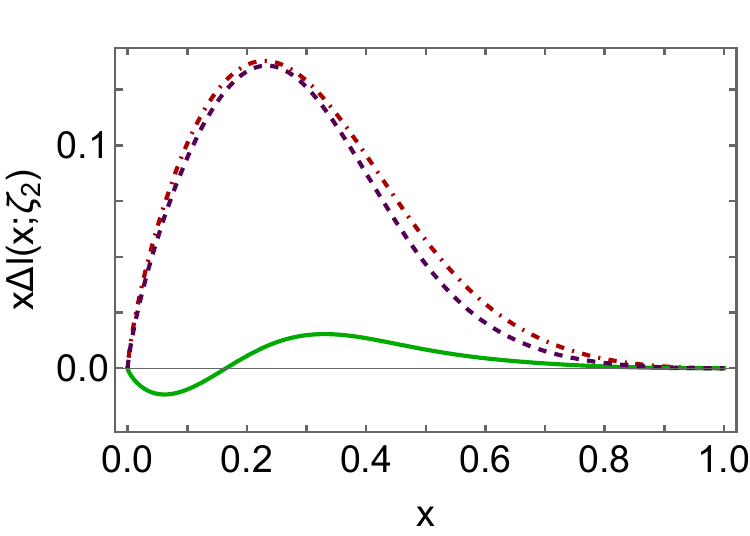}}

\caption{\label{FigPolarisedz2}
Helicity-dependent valence quark DFs in Fig.\,\ref{FigPolarisedzH} evolved to $\zeta=\zeta_2$.
{\sf Panel A}.
$s$ quark in $\Lambda$ -- solid green curve;
$s$ quark in $\Sigma^0$ -- dashed purple curve;
$u$ quark in neutron -- dot-dashed  blue curve.
{\sf Panel B}.
$l$ quark in $\Lambda$ -- solid green curve;
$l$ quark in $\Sigma^0$ -- dashed purple curve;
$0.5\times d$ quark in neutron -- dot-dashed  red curve.
(Neutron results from Ref.\,\cite{Yu:2024qsd}.)
}
\end{figure}

Beginning with the SCI hadron-scale baryon polarised valence-quark DFs in Fig.\,\ref{FigPolarisedzH} and employing the AO evolution scheme \cite{Yin:2023dbw}, one arrives at the $\zeta_2$ polarised valence quark DFs drawn in Fig.\,\ref{FigPolarisedz2}.  Naturally, as explained in connection with Fig.\,\ref{FigPolarisedzH}, the qualitative features of the hadron-scale polarised valence-quark DFs are preserved under evolution.
In this case, the effective large-$x$ exponent is:
\begin{equation}
{\rm HD:} \quad \beta_{\Delta{\rm valence}}^{\rm SCI} \stackrel{0.85 < x < 1}{=} 2.27(3)\,,
\label{largexHDvalence}
\end{equation}
matching Eq.\,\eqref{largexHIvalence} within mutual numerical fitting uncertainties; hence, in line with Eq.\,\eqref{ratioHDHI}.
The values of all axial charges in Eq.\,\eqref{TableMoments} are unchanged under AO evolution.
In Ref.\,\cite[Sects.\,4.2, 4.3]{Yu:2024qsd}, one finds a detailed comparison between predictions for proton helicity-de\-pen\-dent DFs and data.

\begin{figure}[t]
\leftline{\large\sf A}
\vspace*{-3ex}

\centerline{%
\includegraphics[clip, width=0.95\linewidth]{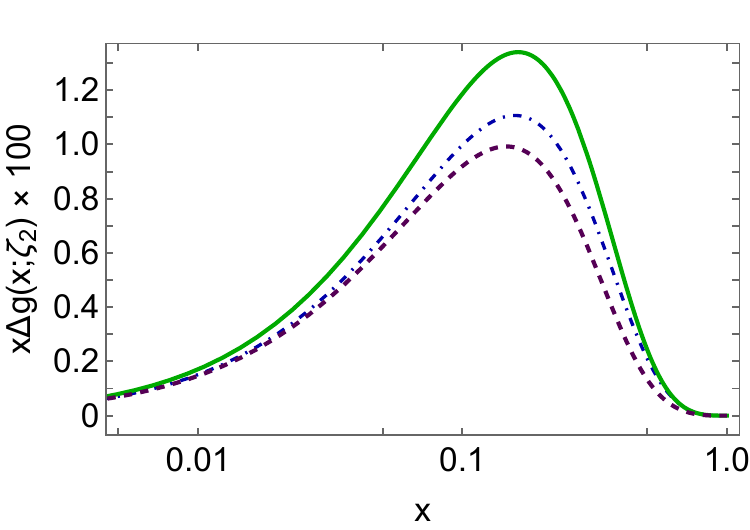}}

\vspace*{1ex}

\leftline{\large\sf B}
\vspace*{-3ex}

\centerline{%
\hspace*{-1ex}\includegraphics[clip, width=0.96\linewidth]{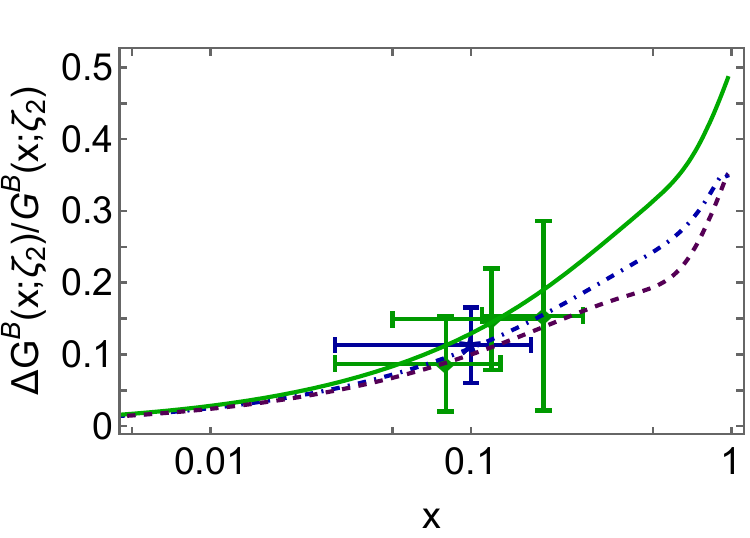}}

\caption{\label{glueHDDFs}
{\sf Panel A}.
Helicity dependent glue DFs at $\zeta=\zeta_2$:
$\Lambda$ -- solid green curve; $\Sigma^0$ -- dashed purple curve; and $p$ -- dot-dashed blue curve.
{\sf Panel B}.
Polarised/unpolarised DF ratio $\Delta G^B(x;\zeta_2)/G^B(x;\zeta_2)$.
$B=\Lambda$ -- solid green curve; $B=\Sigma^0$ -- dashed purple curve; and $B=p$ -- dot-dashed blue curve.
For context, we depict proton values reported in Ref.\,\cite[COMPASS]{COMPASS:2015pim}, which were inferred from measurements at $\zeta=1.73\,$GeV.}
\end{figure}

Nonzero helicity-dependent glue and sea DFs emerge on $\zeta>\zeta_{\cal H}$ following evolution.
The $\zeta=\zeta_2$ glue DFs are drawn in Fig.\,\ref{glueHDDFs}.  They are all non-negative.
Compared with the unpolarised glue DFs, Fig.\,\ref{glueDFs}, the peak magnitudes of the $x$-weighted polarised DFs are roughly an order of magnitude smaller; see Fig.\,\ref{glueHDDFs}\,A.
Evidently, the $\Lambda$ baryon contains more polarised glue than the proton, whereas there is less in the $\Sigma^0$.  Defining
\begin{equation}
\label{DefDeltaG}
\Delta G^B(\zeta) = \int_0^1 dx \Delta G^B(x;\zeta)\,,
\end{equation}
then one finds
\begin{subequations}
\label{DeltaGRatios}
\begin{align}
\Delta G^\Lambda(\zeta_2) /\Delta G^p(\zeta_2) & = 1.13 \,,\\
\Delta G^{\Sigma^0}(\zeta_2) /\Delta G^p(\zeta_2) & = 0.968\,.
\end{align}
\end{subequations}
The effective exponent on the polarised glue DF is:
\begin{equation}
\label{largexHDglue}
{\rm HD:} \quad \beta_{\Delta{\rm glue}}^{\rm SCI} \stackrel{0.85 < x < 1}{=} 3.53(1)\,,
\end{equation}
which, like the exponent on the unpolarised glue DF, is roughly $1+\beta_{\rm valence}^{\rm SCI}$, as in QCD.

SCI predictions for the ratios $\Delta G^B(x;\zeta_{\rm C})/G^B(x;\zeta_{\rm C})$, $B=\Lambda, \Sigma^0, p$, are drawn in Fig.\,\ref{glueHDDFs}\,B.
The uncertainty on available proton data \cite[COMPASS]{COMPASS:2015pim} is too large for any real empirical distinction to be drawn between the different baryons.
This is highlighted by the fact that the mean empirical proton value on the domain covered by measurements is
$ 0.113 \pm 0.038 \pm 0.036$ \cite[COMPASS]{COMPASS:2015pim}.
On the same domain, the SCI predictions for this mean are:
$\Lambda$, $0.159$; $\Sigma^0$, $0.118$; $p$, $0.131$.


\begin{figure}[t]
\centerline{%
\includegraphics[clip, width=0.95\linewidth]{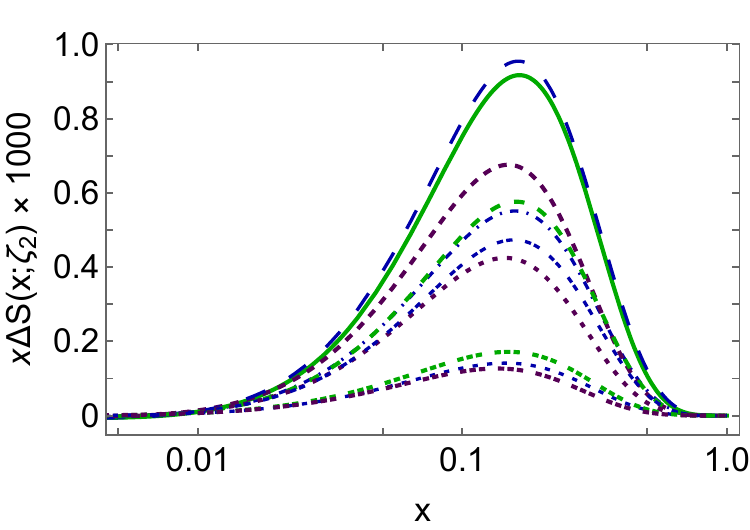}}
\caption{\label{seaHDDFs}
Helicity dependent sea DFs at $\zeta=\zeta_2$.
{\sf Panel A}.
Proton (listed in order of $x\simeq 0.1$ magnitudes).
$2\bar d$  -- long-dashed blue curve;
$2\bar u$  -- dot-dashed blue curve;
$s+\bar s$ -- double-dashed blue curve;
$c+\bar c$ -- short-dashed blue curve.
$\Lambda$ (listed in order of $x\simeq 0.1$ magnitudes).
$2\bar l$  -- solid green curve;
$2\bar s$ -- medium-dashed green curve;
$c+\bar c$ -- triple-dashed green curve.
$\Sigma^0$ (listed in order of $x\simeq 0.1$ magnitudes).
$2\bar l$ -- dashed purple curve;
$2\bar s$  -- short-dashed purple curve;
$c+\bar c$ -- triple-dashed purple curve.
}
\end{figure}

Evolved polarised sea DFs are drawn in Fig.\,\ref{seaHDDFs}.
Compared with unpolarised DFs on the displayed domain, the helicity-dependent DFs are roughly an order of magnitude smaller.
Allowing for the impact of Pauli blocking on the proton light-quark sea, the $\Lambda$-baryon polarised sea $x$-weighted DF profiles have larger peak magnitudes than those in $\Sigma^0, p$.  This is a consequence of the larger polarised glue content of the $\Lambda$.
The effective large-$x$ exponent for SCI polarised sea DFs is
\begin{equation}
{\rm HD:} \quad \beta_{\Delta{\rm sea}}^{\rm SCI} \stackrel{0.85 < x < 1}{=} 4.50(1)\,,
\label{EffectivePowerHDSea}
\end{equation}
which, in keeping with the QCD DGLAP pattern, is roughly $2+\beta_{\rm valence}^{\rm SCI}$.

\section{Baryon Spin}
\label{SecSpin}
Having calculated $\Lambda, \Sigma^0, p$ helicity-dependent DFs, one can address the issue of baryon spin species decomposition.
The singlet axial charges are listed in Eq.\,\eqref{TableMoments}.  At $\zeta_{\cal H}$, glue contributes nothing; so, allowing for the SCI underestimate of $g_A$ by replacing the SCI value with the empirical result, $g_A^{\rm emp}= 1.2754$ \cite{ParticleDataGroup:2024cfk}, the $a_0$ results mean that quark + diquark orbital angular momentum (OAM) carries the following fractions of the baryon spins:
\begin{equation}
\zeta=\zeta_{\cal H} | \quad
\ell_q^{\Lambda} = 29\%\,, \;
\ell_q^{\Sigma^0} = 38.9\%\,, \;
\ell_q^{p} = 36.9\%\,.
\label{qdqL}
\end{equation}

The individual flavour contributions to the light-front orbital angular momentum can be calculated using a so-called Wand\-zura-Wilczek (WW) approximation \cite{Hatta:2012cs, Bhattacharya:2023hbq}:
\begin{equation}
\label{LWW}
\ell_{\mathpzc f}(x;\zeta_{\cal H}) =
x \int_x^1 dy \frac{1}{y^2}
\left[ y{\mathpzc f}(y;\zeta_{\cal H}) - \Delta{\mathpzc f}(y;\zeta_{\cal H})\right]\,,
\end{equation}
where ${\mathpzc f}$, $\Delta{\mathpzc f}$ are the unpolarised and polarised $f$-flavour DFs.
Using Eq.\,\eqref{LWW} and the DFs drawn in Figs.\,\ref{FigUnPolarisedzH}, \ref{FigPolarisedzH}, one obtains the following results for the zeroth moments:
\begin{equation}
\begin{array}{l|ccc}
\zeta_{\cal H}\ & \Lambda\ & \Sigma^0\ & p\ \\\hline
\ell_u & \phantom{-}0.160\phantom{0}\ & 0.00482\ &  0.00018\ \\
\ell_d & \phantom{-}0.160\phantom{0}\ & 0.00482\ & 0.273\phantom{18}\ \\
\ell_s & -0.0753\ & 0.271\phantom{00}\ & \\
\end{array}
\label{qdqLT}
\end{equation}
Evidently, light-quarks carry the bulk of the quark + \!di\-quark OAM in the $\Lambda$-baryon, whereas in $\Sigma^0,p$ the OAM is largely invested in the singly-represented valence quark.  (This means $s$ in $\Lambda, \Sigma^0$ because $l=u=d$.)
These outcomes are a consequence of the diquark correlations in the respective Faddeev wave functions -- see, Eqs.\,\eqref{FaddeevFlavour}, \eqref{protonSF}.
For instance, in $\Lambda$, $s$ quark helicity is dominant because the leading wave function component is $s [ud]$ and angular momentum is stored in the pieces that have $l$-quarks as bystanders to the correlations.
It is worth stressing that, in every case,
\begin{equation}
\zeta = \zeta_{\cal H}| \quad a_0^B/2 + \sum_{q\in B}\ell_q^B = 1/2\,;
\end{equation}
so, the WW formula is a precise measure at the hadron-scale.

Any measurement of a baryon spin works with a conserved current, \emph{viz}.\ a quantity proportional to the zeroth moment of the polarised structure function \linebreak $g_1^B(x;\zeta)$ \cite{Altarelli:1988nr}.  Translated into our notation, using the AO evolution scheme, the relevant observable is:
{\allowdisplaybreaks
\begin{align}
a_{0}^{B{\rm E}}(\zeta) & = a_{0}^B - n_f \frac{\hat \alpha(\zeta)}{2\pi} \Delta G^B(\zeta)\,,
\label{Eqa0E}
\end{align}
where $\Delta G^B(\zeta)$ is given in Eq.\,\eqref{DefDeltaG} and $n_f$ is the number of active quark flavours.
Herein, evolution is defined with $n_f=4$.
Both terms on the right-hand side of Eq.\,\eqref{Eqa0E} are $\zeta$-independent at leading-order in pQCD; hence, in the AO scheme.  So, the observed value of a baryon's helicity may receive a (significant) correction from its gluon helicity despite the presence of the coupling, which runs to zero with increasing $\zeta$ \cite{Altarelli:1988nr}.

Using the SCI results for $\Delta G^B(x;\zeta_{2})$ -- Fig.\,\ref{glueHDDFs} -- one finds, via Eqs.\,\eqref{TableMoments}, \eqref{Eqa0E}:
\begin{equation}
\label{SpinCorrection}
\begin{array}{l|ccc}
\zeta_{2}\ & \ \Lambda\ & \Sigma^0\ & p\ \\\hline
\Delta^B G/g_A\ &\  1.21\phantom{4}\ & 1.04\phantom{4}\ &  1.07\phantom{4}\  \\
a_{0}^{B{\rm E}}/g_A & \ 0.286\  & 0.246\ & 0.254\ \\
\end{array}\,.
\end{equation}
Again correcting for the SCI underestimate of $g_A$, the results in Eq.\,\eqref{SpinCorrection} correspond to
\begin{equation}
a_{0}^{\Lambda{\rm E}} = 0.365\,, \quad
a_{0}^{\Sigma^0{\rm E}} = 0.314\,, \quad
a_{0}^{p{\rm E}} = 0.324\,.
\end{equation}
Regarding the proton, at the scale $\zeta=1.73\,$GeV, the value $0.32(7)$ is reported in Ref.\,\cite[COMPASS]{COMPASS:2016jwv}.  Apparently, contemporary CSM analyses deliver a viable solution of the proton spin crisis, precipitated by the measurements described in Ref.\,\cite{EuropeanMuon:1987isl} -- see, also, Ref.\,\cite[Fig.\,4]{Yin:2023dbw}.

It is worth remarking that Eq.\,\eqref{Eqa0E} has been challenged \cite{Jaffe:1989jz}.
We find grounds for disagreement with counterexamples therein, given the identities and results in Refs.\,\cite{Bhagwat:2007ha, Raya:2016yuj, Ding:2018xwy}.
Nevertheless, we share the view that, owing to the non-Abelian anomaly, it is difficult to supply an unambiguous separation of the proton spin, as measured via the isoscalar axial current, into contributions from quark and gluon partons \cite{Jaffe:1989jz, Bass:2001dg}.
Different resolutions of the proton spin puzzle are possible \cite{Aidala:2012mv, Deur:2018roz}.
In this context, it should be emphasised that we begin with dressed quasiparticle degrees of freedom, not parton-like gluons and quarks, and therefrom deliver results for $a_0^{B{\rm E}}(\zeta_2)$ using AO evolution.
Future analyses will reveal whether the match between our predictions for these quantities and the scale of the COMPASS proton result is accidental or meaningful.

Returning to the question of a light-front separation of baryon spin into contributions from quark and gluon spin and orbital angular momenta (OAM), we observe that the results in Eq.\,\eqref{qdqLT} are scale dependent.   In one common approach to the problem \cite{Jaffe:1989jz}:
\begin{equation}
\label{SpinExpand}
\frac{1}{2} =: \frac{1}{2} a_0^B + \sum_{q\in B} \ell_q^B(\zeta) + \Delta^BG(\zeta) + \ell_g^B(\zeta)\,,
\end{equation}
where the leading quantity is scale invariant, but the remainder evolve.  Taking these things into account,
one obtains the following results: 
\begin{equation}
\label{LSdecompose}
\begin{array}{l|ccc}
\zeta_{2}\ & \ \Lambda\ & \Sigma^0\ & p\ \\\hline
\ell_{u+\bar u}^B & \phantom{-}0.0681\phantom{0}\ & -0.0353\phantom{0}\ &  -0.0967\phantom{77}\ \\
\ell_{d+\bar d}^B & \phantom{-}0.0681\phantom{0}\ & -0.0353\phantom{0}\ & \phantom{-}0.150\phantom{777}\ \\
\ell_{s+\bar s}^B & -0.100\phantom{00}\ &  \phantom{-}0.141\phantom{00}\ &  \phantom{-}0.00995\phantom{0}\ \\
\ell_{c+\bar c}^B & -0.00312\ & \phantom{-}0.00387\  & \phantom{-}0.00380\phantom{0}\  \\
\sum_{q\in B} \ell_q^B\  & \phantom{-}0.0389\phantom{0}\ & \phantom{-}0.0747\phantom{0}\ & \phantom{-}0.0675\phantom{77}\ \\
\Delta^B G\ &\  \phantom{-}1.11\phantom{000}\ & \phantom{-}0.952\phantom{04}\ &  \phantom{-}0.983\phantom{777}\  \\
\ell_g^B &  -0.901\phantom{00}\  & -0.746\phantom{00}\  & -0.777\phantom{777}\  \\\hline
\end{array}
\end{equation}
Here, whilst $\sum_{q\in B} \ell_q^B$ is exact in all cases, the quark flavour separations are estimates based on Eq.\,\eqref{LWW}.  \linebreak When glue contributions are significant, this formula ceases to be exact, delivering instead semiquantitatively reliable results, which we nevertheless find accurate to within roughly 10\%.

Regarding Eq.\,\eqref{SpinExpand} and using the values in Eqs.\,\eqref{SpinCorrection}, \eqref{LSdecompose}
one obtains the following SCI results for the fractional contributions to the baryon spins:
\begin{equation}
\begin{array}{l|ccc}
\zeta_{2}\ & \ \Lambda\ & \Sigma^0\ & p\ \\\hline
{\rm quark\ helicity} & \ 51.0\%\phantom{0}\  & 43.9\%\ & 45.3\%\ \\
{\rm quark\ OAM} & \ \phantom{5}7.79\%\ & 14.9\%\ & 13.5\%\ \\
{\rm gluon\ J} & \ 41.2\%\phantom{0}\ & 41.2\%\ & 41.2\%\ \\\hline
\end{array}
\end{equation}
Using AO evolution, the net glue contribution to each baryon's spin is the same because, overall, it is determined by the quark momentum fraction \cite[Eq.\,(32)]{Yin:2023dbw}, which is the same in each hadron; see Table~\ref{MellinMoments}.
The remainder is lodged with quark helicity and OAM, in the fractions displayed: at $\zeta_2$, quark OAM in the $\Lambda$-baryon is only $\approx 55$\% of that in $\Sigma^0, p$.
Notably, the proton OAM contributions in Eq.\,\eqref{LSdecompose} are semiquantitatively aligned with the lQCD values reported elsewhere \cite{Alexandrou:2020sml}.

\section{Summary and Perspective}
\label{epilogue}
Adopting a quark + interacting-diquark picture of bar\-yon structure, we employed a symmetry-preserving formulation of a vector$\,\times\,$vector contact interaction (SCI) to deliver an extensive, coherent description of helicity-independent and -dependent parton distribution functions (DFs) for $\Lambda, \Sigma^0$ baryons -- valence, glue, and four-flavour separated sea -- and comparisons with kindred nucleon DFs.
$\Lambda, \Sigma^0$ baryons are of special interest because they have the same quark content, $u, d, s$, yet different isospin -- $I=0, 1$, respectively -- and this is expressed in their spin-flavour wave functions.
We assumed isospin symmetry throughout, writing $l=u, d$, but this does not entail SU$(3)$-flavour symmetry.
A \linebreak strength of the SCI is that typical analyses are largely algebraic; hence, the formulae and results are readily understood.  This enables clear judgements to be made of both the SCI outcomes themselves and, via relevant comparisons, results obtained using more sophisticated frameworks.

Working with SCI formulae for had\-ron-scale, $\zeta_{\cal H}$, $\Lambda, \Sigma^0$ valence quark DFs [Sects.\,\ref{HSHIDFs}, \ref{HSHDDFs}], numerical results were presented [Sect.\,\ref{DFresultsHS}].
By definition of $\zeta_{\cal H}$, glue and sea DFs vanish at this scale.
The spin-flavour wave functions of the $\Lambda$, $\Sigma^0$ baryons, which express their diquark content, have an observable impact on the DF predictions.
We anticipate that similar behaviour and conclusions will emerge from quark + diquark approaches to baryon structure that use Schwinger functions which more realistically express basic features of QCD, such as momentum-dependent running quark masses, diquark amplitudes, and baryon Faddeev amplitudes.

Regarding the far-valence domain, for instance, it is only the presence of axialvector diquarks in the $\Sigma^0$ that enables nonzero finite values for $l(x)/s(x)|_{x\simeq 1}$, \linebreak $\Delta l(x)/\Delta s(x)|_{x\simeq 1}$.
Even in the presence of axialvector diquarks, the dominance of scalar diquarks in the $\Sigma^0$ leads to values of these ratios that are large compared with their $\Lambda, p$ analogues [Eq.\,\eqref{FarVx}].
Studying the $\Sigma^0$ spin-flavour amplitude, it becomes clear that only with the existence of axialvector diquark correlations can the $s$ quark participate as a valence degree of freedom in the $\Sigma^0$; otherwise, it is always locked away in an isoscalar-scalar diquark correlation.
Consequently, in the absence of axialvector diquarks, the $s$ quark could carry none of the $\Sigma^0$ spin at $\zeta_{\cal H}$ [Eq.\,\eqref{sDF(Sigma)polarised eq(47)}].

In order to deliver predictions relevant to possible future data, we used the AO scheme \cite{Yin:2023dbw} to evolve the valence-quark DFs to a resolving scale $\zeta  = \zeta_2 := 2\,{\rm GeV} \gg \zeta_{\cal H}$ appropriate to modern experiments \linebreak \, [Sect.\,\ref{EvolvedDFs}].

Regarding valence quark DFs, evolution relocates support to lower $x$; hence, whilst it preserves the hadron-scale patterns of difference and similarity, they steadily become less visible with increasing $\zeta$ as the DF support domain is compressed.
Thus, given known uncertainties in DF inferences from data, our analysis reveals that very precise data and reliable phenomenology would be necessary before the predicted differences between $\Lambda$, $\Sigma^0$, $p$ DFs could be confirmed empirically [Fig.\,\ref{FigUnPolarisedz2}].

Higgs couplings into QCD generate a very large disparity between $l$- and $s$-quark current masses.
However, our analysis revealed that in baryons, as in mesons, this Higgs-driven imbalance is largely masked by the size of emergent hadron mass effects.
Thus, whilst the peak locations in $x$-weighted $\Lambda, \Sigma^0$ valence quark DFs are shifted with respect to those in the nucleon, the relocations are modest and reflect the impacts of both diquark structure and dressed-quark mass differences [Table~\ref{MellinMoments}].

Glue and sea DFs are nonzero $\forall \zeta > \zeta_{\cal H}$.
Regarding helicity-independent glue DFs, the profiles in $\Lambda$, $\Sigma^0$, $p$ are very similar [Fig.\,\ref{glueDFs}].
Differences only become apparent when one considers ratios, which reveal that, on the valence quark domain, both the glue-in-$\Lambda$ and glue-in-$\Sigma^0$ DFs are suppressed compared to the glue-in-$p$ DF, with the in-$\Sigma^0$ suppression being greater.
Helicity independent sea DFs are qualitatively and semiquantitatively similar to the glue DFs [Fig.\,\ref{SeaDFs}].
Notably, our SCI analysis predicts a nonzero $c+\bar c$ DF in each hadron discussed herein, \emph{viz}.\ $\Lambda, \Sigma^0, p$: in each case, they have sea quark profiles and fairly commensurate magnitudes.

Helicity dependent glue and sea DFs also emerge on $\zeta>\zeta_{\cal H}$.
The helicity-dependent glue DF is non-negative in each hadron considered; and compared with kindred unpolarised glue DFs, the peak magnitudes of the $x$-weighted polarised DFs are roughly an order of magnitude smaller [Fig.\,\ref{glueHDDFs}\,A].
The SCI predicts that the $\Lambda$ baryon contains more polarised glue than the proton, whereas there is less in the $\Sigma^0$ [Eq.\,\eqref{DeltaGRatios}].
However, these differences are small.
They could not be distinguished by data whose precision does not greatly exceed that which has thus far been achieved [Fig.\,\ref{glueHDDFs}\,B].
Helicity-dependent $x$-weighted sea DFs are an order-of-magnitude smaller than their unpolarised partners.

With the entire array of DFs available, it was possible to discuss $\Lambda, \Sigma^0, p$ spin decompositions.
At the hadron scale, with glue and sea absent, light-quarks carry most of the quark + di\-quark orbital angular momentum (OAM) in the $\Lambda$-baryon, whereas in $\Sigma^0,p$ the OAM is largely invested in the singly-represented valence quark.

Spin and OAM contributions evolve with scale.
Accounting for the non-Abelian anomaly contribution to the flavour-singlet current, the SCI delivers a viable explanation of proton spin measurements, with predicted values for each hadron lying within contemporary empirical bounds.
In detail, at $\zeta_2$, compared with the proton value, quarks carry 12\% more of the $\Lambda$ spin and $3$\% less of that of the $\Sigma^0$.
The net glue contribution to each baryon's spin is the same, viz.\ $\approx 40$\%;
and regarding quark OAM, that in the $\Lambda$-baryon is only $\approx 55$\% of that in $\Sigma^0, p$.

This study completes a second step in a systematic programme that aims to deliver QCD-connected predictions for all baryon DFs and their unification with those of mesons.
A natural continuation would see this analysis revisited using QCD-connec\-ted Schwinger functions with more realistic momentum dependence.  Overall, we expect qualitative equivalence, but there may be some differences owing to rigidities in SCI propagators and amplitudes, which would be worth exposing.

It would also be useful to develop the framework further so that one could deliver predictions for spin transfer in deep-inelastic $\Lambda$ electroproduction.  Data exists \cite{HERMES:1999buc, Schnell:2024ppe} and more is expected from modern and anticipated facilities, but the lack of sound information on the fragmentation functions involved is an impediment to such an extension \cite{Metz:2016swz}.

Calculations of DFs for decuplet baryons would also be of interest \cite{Zhao:2024zpy} because their spin-flavour amplitudes are much simpler than those for states in the baryon octet \cite{Yin:2021uom, Liu:2022ndb}.  The decuplet-octet contrasts could therefore provide additional empirical signatures of the structural impacts of diquark correlations.
Analyses of DFs characterising baryons containing one or more heavy quarks may also be valuable insofar as, \emph{e.g}., they would expose the impact on EHM expressions made by stronger Higgs boson couplings into QCD.

In the longer term, one may reasonably expect to repeat this and related analyses in calculations that begin with a Poincar\'e-covariant Faddeev equation treatment of the three-valence-body bound state problem.

\begin{CJK*}{UTF8}{gbsn}
\begin{acknowledgements}
We are grateful to Z.-F.\ Cui (崔著钫) for valuable discussions.
Work supported by:
National Natural Science Foundation of China (grant no.\ 12135007);
and
Natural Science Foundation of Anhui Province (grant no.\ 2408085QA028).
\end{acknowledgements}
\end{CJK*}

\begin{small}

\noindent\textbf{Data Availability Statement} Data will be made available on reasonable request.  [Authors' comment: All information necessary to reproduce the results described herein is contained in the material presented above.]
\medskip

\noindent\textbf{Code Availability Statement} Code/software will be made available
on reasonable request. [Authors' comment: No additional remarks.]

\end{small}

\appendix

\section{Diquark Distribution Functions}
\label{diquarkDFs}
Formulae in Sects.\,\ref{HSHIDFs}, \ref{HSHDDFs} involve hadron-scale, $\zeta_{\cal H}$, valence-quark DFs associated with the following diquarks:
$[ud]$, $\{ud\}$,
$[ls]$, $\{ls\}$.
To calculate them, we exploit the fact that so long as the mass-disparity between the two valence degrees-of-freedom in the system is not too large, as is the case for $u, d, s$ systems, then it is a good approximation to write the hadron-scale DF as the square of the related distribution amplitude ($J=0, 1$):
\begin{equation}
{\mathpzc q}_V^{{q q^\prime}_{\!\!J}}(x) =
{\mathpzc n}_{{q q^\prime }_{\!\!J}} \varphi_{{\underline{q}q^\prime }_{\!\!J}}^2(x)\,,
\label{qqDF1}
\end{equation}
with the constant of proportionality, ${\mathpzc n}_{{q q^\prime }_{\!\!J}}$,  chosen to ensure unit normalisation of the DF:
\begin{equation}
\int_0^1 dx\, {\mathpzc q}_V^{{qq^\prime }_{\!\!J}}(x)  = 1\,.
\label{qqDF2}
\end{equation}
The veracity of this approximation is discussed, \emph{e.g}., in Ref.\,\cite[Sect.\,3]{Roberts:2021nhw}

Using Eqs.\eqref{qqDF1}, \eqref{qqDF2}, the first two desired DFs are made available via  Ref.\,\cite[Eqs.\,(19), (24)]{Lu:2021sgg}.

Turning to $[ls]$, $\{ls\}$, the distribution amplitudes (DAs) are defined by the following identities ($N_c^{\bar 3}=2$):
\begin{align}
f_{{[ls]}}\varphi_{{[\underline{l}s]}}&(x) = N^{\bar3}_{c}\textup{tr}_{\textup{D}} \nonumber \\
&\times
\int\frac{d^4 k}{(2\pi)^4} \delta^{xP}_n(k_{\eta})\gamma_{5}{\gamma\cdot n} \chi^{C^{\dagger}}_{{[\underline{l}s]}}(k_{\eta\bar \eta},P), \nonumber\\
\chi^{C^{\dagger}}_{{[\underline{l}s]}}(k_{\eta\bar \eta},P) & =S_l(k_{\eta}) \Gamma^{[\underline{l}s]}(k_{\eta\bar \eta},P)C^{\dagger}S_s(k_{\bar \eta}),
\label{[qs] diquark DA eq54}
\end{align}
where the trace is over spinor indices and $k_{\eta\bar\eta} = [k_\eta+k_{\bar\eta}]/2$, $k_\eta  = k + \eta P$, $k_{\bar\eta}  = k - (1-\eta) P$.
Similarly,
\begin{align}
{n\cdot P} f_{{\{ls\}}} & \varphi_{{\{\underline{l} s\}}}(x) = m_{{\{l s\}}} N^{\bar3}_{c}\textup{tr}_{\textup{D}} \nonumber \\
& \times \int \frac{d^4 k}{(2\pi)^4} \delta^x_{nP} (k_{\eta}){\gamma\cdot n} \, n\cdot \chi^{C^{\dagger}}_{{\{\underline{l}s\}}}(k_{\eta\bar \eta},P),\nonumber \\
\chi^{C^{\dagger}}_{{\{\underline{l}s\}}\nu}(k_{\eta\bar \eta},P)&=S_l(k_{\eta})\Gamma^{\{\underline{l}s\}}_\nu(k_{\eta\bar \eta},P)C^{\dagger} S_s(k_{\bar \eta})\,.
\label{{qs} diquark DA eq55}
\end{align}

These expressions are readily evaluated using standard SCI regularisation procedures and the masses and amplitudes specified by Eq.\,\eqref{qqBSAs} and Table~\ref{qqBSAsolutions}.  The ``decay constants'',
$f_{{[ls]}}$, $f_{{\{ls\}}}$, are simply mass-dimension one quantities defined by the requirement
\begin{equation}
\int_0^1 dx\, \varphi_{{\underline{q}q^\prime }_{\!\!J}}^2(x) = 1\,.
\end{equation}
Given Eqs.\eqref{qqDF1}, \eqref{qqDF2}, their values are immaterial.

Using Eqs.\,\eqref{[qs] diquark DA eq54}, \eqref{{qs} diquark DA eq55}, one may readily verify the following symmetry relations:
\begin{subequations}
\label{DAsymmetry}
\begin{align}
\varphi_{{[\underline{l}s]}}(x) & = \varphi_{{[l\underline{s}]}}(1-x)\,,\\
\varphi_{\{{\underline{l}s\}}}(x) & = \varphi_{\{l\underline{s}\}}(1-x)\,.
\end{align}
\end{subequations}

Completing the calculations defined above, one arrives at the following DFs:
\begin{subequations}
\label{qqDFs}
\begin{align}
{\mathpzc l}_V^{[ls]}(x) & =
{\mathpzc n}_{[ls]} \big\{
F_{{[ls]}}{\cal C}^{\rm iu}_{0}(\sigma_{[ls]})\notag\\
& +  [ \sigma_{[ls]}F_{{[ls]}} + M_l ( 2 M_{ls} E_{{[ls]}} - [M_l+M_s] F_{{[ls]}})\nonumber \\
& + x(M_l^2-M_s^2) F_{[ls]} \nonumber\\
& + 2x M_{ls}(M_s-M_l) E_{[ls]}] \overline{\cal C}^{\rm iu}_1(\sigma_{[ls]})
\big\}^2\,,
\label{SCI of [qs] diquark DF eq58}  \\
{\mathpzc l}_V^{\{ls\}}(x) & = n_{\{ls\}} E^2_{\{ls\}}
\big\{ {\cal C}^{\rm iu}_{0}(\sigma_{\{ls\}})
+ \big[\sigma_{{\{ls\}}}
\nonumber\\
& + 2 x (1-x) m^2_{{\{ls\}}} + x(M_l^2-M_s^2)  \nonumber \\
& + M_l (M_s - M_l)\big] \overline{\cal C}^{\rm iu}_1(\sigma_{{\{ls\}}})
\big\}^2\,,
\label{SCI of {qs} diquark DF eq59}
\end{align}
\end{subequations}
where $\sigma_{[ls]}=(1-x)M_l^2+xM_s^2-x(1-x){m^2_{[ls]}}$, $\sigma_{{\{ls\}}}=(1-x)M_l^2+xM_s^2-x(1-x){m^2_{{\{ls\}}}}$.
Equations~\eqref{DAsymmetry} entail the hadron-scale identities:
\begin{subequations}
\label{DFsymmetry}
\begin{align}
{\mathpzc s}_V^{{[ls]}}(x)& = {\mathpzc l}_V^{{[ls]}}(1-x)\,, \\
{\mathpzc s}_V^{{\{ls\}}}(x) & = {\mathpzc l}_V^{{\{ls\}}}(1-x)\,.
\end{align}
\end{subequations}

It only remains to specify the $[ls] \leftrightarrow \{ls\}$ transition DFs.  This has not yet been calculated.
The proton study of Ref.\,\cite{Yu:2024qsd} required the $[ud] \leftrightarrow \{ud\}$ analogue.
It employed the scale free DF $30 x^2(1-x)^2$ and two other forms.  There was little sensitivity to the choice; see, Ref.\,\cite[Fig.\,2]{Yu:2024qsd}.
Exploiting that outcome, we use a single form, based on the ``middle'' kaon DA described in Ref.\,\cite[Eq.\,(45), Table~3]{Cui:2020tdf}:
\begin{align}
\varphi_K^l(x) & \propto x(1-x) \nonumber \\
& \big[
1 + \rho x^{\alpha/2}(1-x)^{\beta/2} + \gamma x^\alpha (1-x)^\beta
\big]\,,
\end{align}
$\rho = 5.00$, $\gamma = -5.97$, $\alpha=0.0638$, $\beta = 0.0481$, and write
\begin{align}
{\mathpzc l}_V^{01}(x) & = {\mathpzc n}_{01} [\varphi_K^l(x)]^2 =
{\mathpzc s}_V^{01}(1-x) \,,
\end{align}
with ${\mathpzc n}_{01} $ ensuring unit normalisation, as usual.

\section{Baryon Distribution Functions}
\label{DFappendix}
%
In this appendix, all formulae refer to DFs at the hadron scale, $\zeta_{\cal H}$.
They use the
masses in Tables~\ref{Tab:DressedQuarks}, \ref{qqBSAsolutions}
and the
coefficients describing the Faddeev equation solutions that are listed in Table~\ref{FadSolution}.

\subsection{$\Lambda$ -- helicity independent}
\label{AppLambdaHI}
In order to obtain numerical results for each of the terms in Eqs.\,\eqref{sDF(Lambda) eq(2)}, \eqref{uDF(Lambda) eq(9)}, it is sufficient to use the following formulae, obtained via standard SCI regularisation procedures, isospin symmetry, and
Eqs.\,\eqref{sDF2and3(Lambda) eq(5)}, \eqref{sDF4and5(Lambda) eq(7)}, Eqs.\,\eqref{ExchangeLambda}.

{\allowdisplaybreaks
\begin{subequations}
\begin{align}
&{\mathpzc s}_{V_{Q^{[ud]}}}^{\Lambda}  (x)  =
 [{\mathpzc s}^{r_1^0}]^2 (1-x) \big[\, \overline{\cal C}^{\rm iu}_1(\omega_{[ud]}) \nonumber \\
 &
 \qquad + 2 x (2 M_s m_{\Lambda} - {\mathpzc t}_{[ud]} ) \bar{\cal C}^{\rm iu}_2(\omega_{[ud]})/ \omega_{[ud]}
 \big] , \label{s_{V_{Q^[ud]_0}} eq(63a)} \\
%
&{\mathpzc u}_{V_{Q^{[ls]}}}^{\Lambda}  (x)  =
 [{\mathpzc s}^{{\mathpzc r}_2^0}]^2 (1-x) \big[ \,\overline{\cal C}^{\rm iu}_1(\omega_{[ls]})  \nonumber \\
 &
\qquad + 2 x (2 M_l m_{\Lambda} - {\mathpzc t}_{[ls]} ) \bar{\cal C}^{\rm iu}_2(\omega_{[ls]})/ \omega_{[ls]}
 \big] , \label{u_{V_{Q^[ds]_0}} eq(63b)} \\
%
&{\mathpzc u}_{V_{Q^{\{ls\}}}}^{\Lambda}  (x)  = {\mathpzc u}_{11}^{\Lambda}  (x) +
{\mathpzc u}_{12}^{\Lambda}  (x) +
{\mathpzc u}_{22}^{\Lambda}  (x) \,, \label{u_{V_{Q^{ds}_0}} eq(61c)} \\
%
&{\mathpzc u}_{11}^{\Lambda}  (x)  =
2 [{\mathpzc a}_{1}^{r_3^0}]^2 (1-x) \big[ \,\overline{\cal C}^{\rm iu}_1(\omega_{\{ls\}})
\nonumber \\
 &
 \qquad + 2 x (4 M_l m_{\Lambda} - {\mathpzc t}_{\{ls\}} ) \bar{\cal C}^{\rm iu}_2(\omega_{\{ls\}})/ \omega_{\{ls\}}
 \big] , \label{u_{11} eq(61d)} \\
%
&{\mathpzc u}_{12}^{\Lambda}  (x)  =
- 2 {\mathpzc a}_{1}^{r_3^0}{\mathpzc a}_{2}^{r_3^0}
 (1-x) \big[\,\overline{\cal C}^{\rm iu}_1(\omega_{\{ls\}})
 \nonumber \\
 & \qquad
  - 2 x (2 M_l m_{\Lambda} + {\mathpzc t}_{\{ls\}} ) \overline{\cal C}^{\rm iu}_2(\omega_{\{ls\}})/ \omega_{\{ls\}}
 \big] , \label{u_{12} eq(61e)} \\
%
&{\mathpzc u}_{22}^{\Lambda}  (x)  =
- 
[{\mathpzc a}_{2}^{r_3^0}]^2 (1-x) \big[ \,\overline{\cal C}^{\rm iu}_1(\omega_{\{ls\}})  \nonumber \\
 & \qquad
 - 2 x (2 M_l m_{\Lambda} + {\mathpzc t}_{\{ls\}} ) \overline{\cal C}^{\rm iu}_2(\omega_{\{ls\}})/ \omega_{\{ls\}}
 \big] , \label{u_{22} eq(61f)}
\end{align}
\end{subequations}
where
\begin{subequations}
\begin{align}
\omega_{[ud]} & = x m_{[ud]}^2 + (1-x) M_s^2 - x(1-x) m_{\Lambda}^2 \, ,\label{omega_[ud] eq(62a)}\\
\omega_{[ls]} & = x m_{[ls]}^2 + (1-x) M_l^2 - x(1-x) m_{\Lambda}^2 \, ,\label{omega_[qs] eq(62b)}\\
\omega_{\{ls\}} & = x m_{\{ls\}}^2 + (1-x) M_l^2 - x(1-x) m_{\Lambda}^2 \, ,\label{omega_{qs} eq(62c)}\\
{\mathpzc t}_{[ud]} & = m_{[ud]}^2 - M_s^2 -m_{\Lambda}^2 \, ,\label{t_[ud] eq(62d)}\\
{\mathpzc t}_{[ls]} & = m_{[ls]}^2 - M_l^2 -m_{\Lambda}^2 \, ,\label{t_[qs] eq(62e)}\\
{\mathpzc t}_{\{ls\}} & = m_{\{ls\}}^2 - M_l^2 -m_{\Lambda}^2.\label{t_{qs} eq(62f)}
\end{align}
\end{subequations}
Each right-hand side should be multiplied by $1/[16\pi^2]$.
}

\subsection{$\Sigma^0$ -- helicity independent}
In order to calculate Eqs.\,\eqref{sDF(Sigma) eq(23)}, \eqref{uDF(Sigma)  eq(26)}, it is sufficient to use the following formulae, Eq.\,\eqref{sVQfDud}, and relevant expressions in \ref{AppLambdaHI} as directed in Sect.\,\ref{SSSigma0HI}.
{\allowdisplaybreaks
\begin{subequations}
\begin{align}
&{\mathpzc u}_{V_{Q^{[ls]}}}^{\Sigma}  (x)
={\mathpzc u}_{V_{Q^{[ls]}}}^{\Lambda}(x)|_{m_{\Lambda}\to m_{\Sigma}}^{r_2^0\to r_1^1}, \label{u_{V_{Q^{ds}_0}}(Sigma) eq(63a)} \\
%
&{\mathpzc u}_{V_{Q^{\{ls\}}}}^{\Sigma}  (x)
={\mathpzc u}_{V_{Q^{\{ls\}}}}^{\Lambda}(x)|_{m_{\Lambda}\to m_{\Sigma}}^{r_3^0\to r_3^1}, \label{u_{V_{Q^{ds}_0}}(Sigma) eq(63b)} \\
%
&{\mathpzc s}_{V_{Q^{\{ud\}}}}^{\Sigma}  (x)  =  [{\mathpzc s}_{11}^{\Sigma}  (x) +
{\mathpzc s}_{12}^{\Sigma}  (x) +
{\mathpzc s}_{22}^{\Sigma}  (x) ]/(4\pi)^2
\,, \label{s_{V_{Q^{ud}_0}} eq(63c)}\\
%
&{\mathpzc s}_{11}^{\Sigma}  (x)  =
2 [{\mathpzc a}_{1}^{r_2^1}]^2 (1-x)
\big[\, \overline{\cal C}^{\rm iu}_1(\omega_{\{ud\}}) \nonumber \\
 &
\qquad  + 2 x (4 M_s m_{\Sigma} - {\mathpzc t}_{\{ud\}} ) \overline{\cal C}^{\rm iu}_2(\omega_{\{ud\}})/ \omega_{\{ud\}} \big] , \label{s_{11} eq(63d)} \\
%
&{\mathpzc s}_{12}^{\Sigma}  (x)  =
-2 {\mathpzc a}_{1}^{r_2^1} {\mathpzc a}_{2}^{r_2^1} (1-x) \big[\,
\overline{\cal C}^{\rm iu}_1(\omega_{\{ud\}})  \nonumber \\
 &
\qquad  - 2 x (2 M_s m_{\Sigma} + {\mathpzc t}_{\{ud\}} ) \bar{\cal C}^{\rm iu}_2(\omega_{\{ud\}})/ \omega_{\{ud\}}
 \big] , \label{s_{12} eq(63e)} \\
%
&{\mathpzc s}_{22}^{\Sigma}  (x)  =
- [{\mathpzc a}_{2}^{r_2^1}]^2 (1-x)\big[\, \overline{\cal C}^{\rm iu}_1(\omega_{\{ud\}})
 \nonumber \\
 & \qquad
 - 2 x (2 M_s m_{\Sigma} + {\mathpzc t}_{\{ud\}} ) \overline{\cal C}^{\rm iu}_2(\omega_{\{ud\}})/ \omega_{\{ud\}} \big]
\, , \label{s_{22} eq(63f)}
\end{align}
\end{subequations}
where
\begin{subequations}
\begin{align}
\omega_{\{ud\}} & = x m_{\{ud\}}^2 + (1-x) M_s^2 - x(1-x) m_{\Sigma}^2 \, ,\label{omega_{ud} eq(64a)}\\
{\mathpzc t}_{\{ud\}} & = m_{\{ud\}}^2 - M_s^2 -m_{\Sigma}^2.\label{t_{ud} eq(64b)}
\end{align}
\end{subequations}
}

\subsection{$\Lambda$ -- helicity dependent}
Working from Eqs.\,\eqref{sDF(Lambda)polarised eq(32)}, \eqref{uDF(Lambda)polarised eq(39)}, the following new formulae arise, which should be evaluated using appropriate values from Tables~\ref{Tab:DressedQuarks}\,--\,\ref{FadSolution}.
{\allowdisplaybreaks
\begin{subequations}
\label{LambdPolApp}
\begin{align}
\Delta & {\mathpzc s}_{V_{Q^{[ud]}}}^{\Lambda}(x)  =
 [{\mathpzc s}^{r_1^0}]^2 {\mathpzc A}_s (1-x)  \big[
 -\overline{\cal C}^{\rm iu}_1(\omega_{[ud]}) \nonumber \\
 & \qquad +
2  (2 x  M_s m_{\Lambda} + \tilde{\mathpzc t}_{[ud]} ) \overline{\cal C}^{\rm iu}_2(\omega_{[ud]})/ \omega_{[ud]}\big]\, , \label{Delta s_{V_{Q^[ud]_0}} eq(65a)} \\
%
\Delta & {\mathpzc u}_{V_{Q^{[ls]}}}^{\Lambda}(x)  =
 [{\mathpzc s}^{r_2^0}]^2 {\mathpzc A}_l (1-x)  \big[ -\overline{\cal C}^{\rm iu}_1(\omega_{[ls]})
 \nonumber \\
 & \qquad + 2  (2 x  M_l m_{\Lambda} + \tilde{\mathpzc t}_{[ls]} ) \overline{\cal C}^{\rm iu}_2(\omega_{[ls]})/ \omega_{[ls]} \big]\, , \label{Delta u_{V_{Q^[ds]_0}} eq(65b)} \\
%
\Delta & {\mathpzc u}_{V_{Q^{\{ls\}}}}^{\Lambda} (x)  = \Delta{\mathpzc u}_{11}^{\Lambda} (x) +
\Delta{\mathpzc u}_{12}^{\Lambda} (x) +
\Delta{\mathpzc u}_{22}^{\Lambda} (x) \,,\label{Delta u_{V_{Q^{ds}_0}} eq(65c)} \\
%
\Delta & {\mathpzc u}_{11}^{\Lambda} (x)  =
2 [{\mathpzc a}_{1}^{r_3^0}]^2 {\mathpzc A}_l (1-x)
\big[ \overline{\cal C}^{\rm iu}_1(\omega_{\{ls\}}) \nonumber \\
 & \qquad - 2 \tilde{\mathpzc t}_{\{ls\}} \overline{\cal C}^{\rm iu}_2(\omega_{\{ls\}})/ \omega_{\{ls\}}
 \big] \,, \label{Delta u_{11} eq(65d)} \\
%
\Delta & {\mathpzc u}_{12}^{\Lambda} (x)  =
2 {\mathpzc a}_{1}^{r_3^0} {\mathpzc a}_{2}^{r_3^0} {\mathpzc A}_l (1-x)
\big[ \overline{\cal C}^{\rm iu}_1(\omega_{\{ls\}})   \nonumber \\
 &
\qquad + 2 (2 x M_l m_{\Lambda} - \tilde {\mathpzc t}_{\{ls\}} ) \overline{\cal C}^{\rm iu}_2(\omega_{\{ls\}})/ \omega_{\{ls\}}
 \big] \,, \label{Delta u_{12} eq(65e)} \\
%
\Delta & {\mathpzc u}_{22}^{\Lambda} (x)  =
[{\mathpzc a}_{2}^{r_3^0}]^2 {\mathpzc A}_l (1-x) \big[ \overline{\cal C}^{\rm iu}_1(\omega_{\{ls\}})
\nonumber \\
 & \qquad
 + 2 (2 x M_l m_{\Lambda} - \tilde {\mathpzc t}_{\{ls\}} ) \bar{\cal C}^{\rm iu}_2(\omega_{\{ls\}})/ \omega_{\{ls\}}
 \big] \,, \label{Delta u_{22} eq(65f)}
\end{align}
\end{subequations}
where ${\mathpzc A}_{s,l}$ are given after Eqs.\,\eqref{Aafterhere1}, \eqref{T1uHDL}, and
\begin{subequations}
\begin{align}
\tilde {\mathpzc t}_{[ud]} & = x m_{[ud]}^2 +(2-x) M_s^2 - x(1-2x)m_{\Lambda}^2 \, , \label{tilde t_[ud] eq(66a)}\\
\tilde {\mathpzc t}_{[ls]} & = x m_{[ls]}^2 +(2-x) M_l^2 - x(1-2x)m_{\Lambda}^2 \, , \label{tilde t_[ls] eq(66b)}\\
\tilde {\mathpzc t}_{\{ls\}} & = x m_{\{ls\}}^2 +(2-x) M_l^2 - x(1-2x)m_{\Lambda}^2 \, . \label{tilde t_{ls} eq(66c)}
\end{align}
\end{subequations}
In Eqs.\,\eqref{Delta s_{V_{Q^[ud]_0}} eq(65a)}\,--\,\eqref{Delta u_{V_{Q^{ds}_0}} eq(65c)}, the right-hand sides should be multiplied by $1/(4\pi)^2$.

For the calculation of
$\Delta{\mathpzc f}_{D^{\{ls\}}}^{\Lambda}(x)$
and
$\Delta{\mathpzc f}_{01}^{\Lambda}(x)$
in Sect.\,\ref{LambdaForm},
we use Eqs.\,(37b), (39b) and Table~XIV from Ref.\,\cite{Cheng:2022jxe} to obtain the following:
\begin{subequations}
\begin{align}
\Gamma_{5\mu;\alpha\beta}^{AAls}(\ell,\ell) & = \phantom{-}\frac{1}{2}(0.492+0.564)\, \varepsilon_{\mu\alpha\beta\delta} 2 \ell_\delta \,, \label{Gamma^{AAqs} eq(67a)}\\
\Gamma_{5\mu;\alpha}^{SAls}(\ell,\ell) & = - \frac{1}{2}(0.641+0.742)\,  [m_{[ls]}+m_{\{ls\}}] i \delta_{\mu\alpha}\,. \label{Gamma^{SAqs} eq(67b)}
\end{align}
\end{subequations}
Note that here, for consistency with Eq.\,\eqref{AVdqprop}, we have replaced the transverse momentum projection operators in the relevant Ref.\,\cite{Cheng:2022jxe} formulae by a Kronecker-$\delta$.
The factor of $1/2$ is needed because Ref.\,\cite{Cheng:2022jxe} reported the flavour unseparated $Q^2=0$ value of the form factors, whereas we need the individual components; and we use $1/2$ for simplicity because the variation from the mean is just 7\%, so a detailed flavour separation delivers no material gain.

Proceeding as described, one obtains:
\begin{subequations}  
\label{sterms}
\begin{align}
\Delta {\mathpzc f}_{D^{\{ls\}}}^{\Lambda}(x)  & =
1.056\,
[{\mathpzc a}_{1}^{r_3^0}]^2 x \big\{
\overline{\cal C}^{\rm iu}_1(\tilde\omega_{\{ls\}})
\nonumber \\
& \quad
+ \big[ (M_l +(1-x) m_\Lambda)^2
 - \tilde\omega_{\{ls\}}
\big] \nonumber \\
& \qquad \times 2 \overline{\cal C}^{\rm iu}_2(\tilde\omega_{\{ls\}})/\tilde\omega_{\{ls\}}
\big\}\,, \label{Delta f_{D^{ls}_1} eq(68a)} \\
%
\Delta {\mathpzc f}_{01}^{\Lambda}(x)  & =
(2\times 1.383)\, [m_{[ls]}+m_{\{ls\}}]  \nonumber \\
& \quad \times
{\mathpzc s}^{r_2^0} {\mathpzc a}_{1}^{r_3^0} (1-x) (M_l+[1-x]m_{\Lambda}) \nonumber \\
& \qquad \times \int_0^1 dy \, \overline{\cal C}^{\rm iu}_2(\omega_{01})/\omega_{01} \,, \label{Delta f_{0^+1^+} eq(68b)}
\end{align}
\end{subequations}
where each right-hand side should be multiplied by \linebreak $1/(4\pi)^2$,
$\tilde\omega_{\{ls\}} = \left.\omega_{\{ls\}}\right|_{x\to (1-x)}$,
and
\begin{align}
\omega_{01} & = x M_l^2 + (1-y)(1-x)m_{[ls]}^2 \nonumber \\
& \quad + y (1-x) m_{\{ls\}}^2 -x (1-x) m_{\Lambda}^2\,. \label{omega_{01} eq(69)}
\end{align}
For simplicity, pieces that make practically no numerical contribution to the final results are omitted from Eqs.\,\eqref{sterms}.
}

\subsection{$\Sigma^0$ -- helicity dependent}
Turning finally to Eqs.\,\eqref{sDF(Sigma)polarised eq(47)}, \eqref{uDF(Sigma)polarised eq(50)}, one is led to the following array of results, which again should be evaluated using relevant values from Tables~\ref{Tab:DressedQuarks}\,--\,\ref{FadSolution}.

{\allowdisplaybreaks
\begin{subequations}
\begin{align}
\Delta {\mathpzc u}_{V_{Q^{[ls]}}}^{\Sigma}(x)  &
= \Delta {\mathpzc u}_{V_{Q^{[ls]}}}^{\Lambda}(x)|_{m_{\Lambda}\to m_{\Sigma}}^{r_2^0\to r_1^1}\,, \label{Delta u_{V_{Q^{ds}_0}}(Sigma) eq(70a)} \\
%
\Delta {\mathpzc u}_{V_{Q^{\{ls\}}}}^{\Sigma}  (x)  & = \Delta {\mathpzc u}_{V_{Q^{\{ls\}}}}^{\Lambda}(x)|_{m_{\Lambda}\to m_{\Sigma}}^{r_3^0\to r_3^1}
\,,
\label{Delta u_{V_{Q^{ds}_0}}(Sigma) eq(70b)} \\
\Delta {\mathpzc f}_{D^{\{ls\}}}^{\Sigma} (x) & =  \Delta {\mathpzc f}_{D^{\{ls\}}}^{\Lambda}(x)|_{m_{\Lambda}\to m_{\Sigma}}^{r_3^0\to r_3^1}
\,, \label{Delta f_{D^{qs}_1}(Sigma) eq(70c)} \\
%
\Delta {\mathpzc f}_{01}^{\Sigma} (x)   & =
\Delta {\mathpzc f}_{01}^{\Lambda}(x)|_{m_{\Lambda}\to m_{\Sigma}}^{r_2^0\to r_1^1, r_3^0\to r_3^1}
\,, \label{Delta f_{0^+1^+}(Sigma) eq(70d)} \\
%
\Delta {\mathpzc s}_{V_{Q^{\{ud\}}}}^{\Sigma} (x)   & =
\Delta {\mathpzc s}_{11}^{\Sigma}  (x) +
\Delta {\mathpzc s}_{12}^{\Sigma}  (x) +
\Delta {\mathpzc s}_{22}^{\Sigma}  (x) \,,\label{Delta s_{V_{Q^{ud}_0}} eq(70e)} \\
%
\Delta  {\mathpzc s}_{11}^{\Sigma} (x)  & =
2 [{\mathpzc a}_{1}^{r_2^1}]^2 {\mathpzc A}_s (1-x)
\big[\, \overline{\cal C}^{\rm iu}_1(\omega_{\{ud\}})   \nonumber \\
 & \quad
- 2 \tilde{\mathpzc t}_{\{ud\}} \overline{\cal C}^{\rm iu}_2(\omega_{\{ud\}})/ \omega_{\{ud\}}
 \big] \,, \label{Delta s11 eq(70f)} \\
%
\Delta  {\mathpzc s}_{12}^{\Sigma} (x)   & =
2 {\mathpzc a}_{1}^{r_2^1} {\mathpzc a}_{2}^{{r_2^1}} {\mathpzc A}_s (1-x)
\big[\, \overline{\cal C}^{\rm iu}_1(\omega_{\{ud\}})  \nonumber \\
 &
\quad + 2 (2 x M_s m_{\Sigma} - \tilde {\mathpzc t}_{\{ud\}} ) \nonumber \\
& \qquad \times  \overline{\cal C}^{\rm iu}_2(\omega_{\{ud\}})/ \omega_{\{ud\}} \big]\, , \label{Delta s12 eq(70g)} \\
%
\Delta  {\mathpzc s}_{22}^{\Sigma} (x)  & =
 [{\mathpzc a}_{2}^{r_2^1}]^2 {\mathpzc A}_s (1-x) \big[ \,
 \overline{\cal C}^{\rm iu}_1(\omega_{\{ud\}})
 \nonumber \\
 & \quad
+ 2 (2 x M_s m_{\Sigma} - \tilde {\mathpzc t}_{\{ud\}} ) \nonumber \\
& \qquad \times  \overline{\cal C}^{\rm iu}_2(\omega_{\{ud\}})/ \omega_{\{ud\}}
 \big] \,, \label{Delta s12 eq(70h)}
\end{align}
\end{subequations}
where ${\mathpzc A}_s$ is given after Eq.\,\eqref{Aafterhere1}, and
\begin{align}
\Gamma_{5\mu;\alpha\beta}^{AAud}(\ell,\ell) &  = 0.47 \, \varepsilon_{\mu\alpha\beta\delta} 2 \ell_\delta \,, \label{Gamma^{AAud} eq(71a)}
\end{align}
in Eq.\,\eqref{uDF3(Sigma)polarised eq(51)}, so
\begin{align}
\Delta {\mathpzc f}_{D^{\{ud\}}}^{\Sigma}(x) & = 0.94\, [{\mathpzc a}_{1}^{r_2^1}]^2 x
\big\{\,
\overline{\cal C}^{\rm iu}_1(\tilde\omega_{\{ud\}})  \nonumber \\
&
\quad + \big[ (M_s+[1-x] m_{\Sigma})^2 - \tilde\omega_{\{ud\}}
\big] \nonumber \\
&
\qquad \times
2 \bar{\cal C}^{\rm iu}_2(\tilde\omega_{\{ud\}})/\tilde\omega_{\{ud\}}
\big\}\,. \label{Delta f_{D^{ud}_1} eq(71b)}
\end{align}
The right-hand sides of Eqs.\,\eqref{Delta s_{V_{Q^{ud}_0}} eq(70e)}, \eqref{Delta f_{D^{ud}_1} eq(71b)} should be multiplied by $1/(4\pi)^2$,
$\tilde\omega_{\{ud\}} = \left.\omega_{\{ud\}}\right|_{x\to (1-x)}$, and
\begin{align}
\tilde {\mathpzc t}_{\{ud\}} & = x m_{\{ud\}}^2 +(2-x) M_s^2 - x(1-2x)m_{\Sigma}^2 \, .
\label{tilde t_{ud} eq(72)}
\end{align}
}

\section{Distribution Function Interpolation Coefficients}
\label{DFinterpolationappendix}
All SCI baryon DFs computed herein can reliably be reproduced using the function in Eq.\,\eqref{InterpolateDFs} with the interpolation coefficients listed in Tables~\ref{IcoeffsLambda}, \ref{IcoeffsSigma}.

\begin{table}[t]
\caption{\label{IcoeffsLambda}
Interpolation coefficients for hadron scale in-$\Lambda$ DFs, to be used in Eq.\,\eqref{InterpolateDFs}.
Each entry should be divided by $10^{4}$.
}
\begin{tabular*}
{\hsize}
{
l@{\extracolsep{0ptplus1fil}}
c@{\extracolsep{0ptplus1fil}}
c@{\extracolsep{0ptplus1fil}}
c@{\extracolsep{0ptplus1fil}}
c@{\extracolsep{0ptplus1fil}}}\hline
${\mathpzc f}$ & ${\mathpzc l}$ & ${\mathpzc s}$ & $\Delta{\mathpzc u}$ & $\Delta{\mathpzc s}$ \\\hline
${\mathpzc n}_{\mathpzc f}$  & $\phantom{-}3031\ $  & $\phantom{-}3938\ $  & $\phantom{-}76.49\phantom{0}\ $ & $\phantom{-}2052\ $  \\
$a_1^{\mathpzc f}$  & $\phantom{-}3393\ $ & $\phantom{-}995.5\ $ & $-9769\phantom{0}\ $ & $\phantom{-}1079\ $  \\
$a_2^{\mathpzc f}$  & $-1310\ $  & $-2607\ $ & $-10520\ $ & $-2318\ $   \\
$a_3^{\mathpzc f}$  & $-1262\ $ & $-655.8\ $ & $-2210\phantom{0}\ $ & $-481.7\ $   \\
$a_4^{\mathpzc f}$  & $-106.4\ $ & $\phantom{-}649.9\ $ & $\phantom{-}2911\phantom{0}\ $ & $\phantom{-}612.8\ $   \\
$a_5^{\mathpzc f}$  & $\phantom{-}229.9\ $ & $\phantom{0}262.5$ & $\phantom{-}2546\phantom{0}\ $ & $\phantom{-}144.1\ $   \\
$a_6^{\mathpzc f}$  & $\phantom{-}89.14\ $ & $-137.7\ $ & $\phantom{0}268.2\phantom{0}\ $ & $-181.3\ $  \\
$a_7^{\mathpzc f}$  & $-16.64\ $ & $-76.95\ $  & $-597.3\phantom{0}\ $ & $-54.97\ $  \\
$a_8^{\mathpzc f}$  & $-19.45\ $ & $\phantom{-}23.21\ $ & $-193.3\phantom{0}\ $ & $\phantom{-}37.73\ $  \\
$a_9^{\mathpzc f}$  & $-2.005\ $ & $\phantom{-}13.98\ $ &$\phantom{-}125.6\phantom{0}\ $  & $\phantom{-}10.91\ $   \\
$a_{10}^{\mathpzc f}$  & $\phantom{-}1.452\ $ & $-2.519\ $ & $\phantom{-}73.69\phantom{0}\ $ & $-5.356\ $ \\\hline
\end{tabular*}
\end{table}

\begin{table}[t]
\caption{\label{IcoeffsSigma}
Interpolation coefficients for $\Sigma^0$ hadron scale DFs, to be used in Eq.\,\eqref{InterpolateDFs}.
Each entry should be divided by $10^{4}$.
}
\begin{tabular*}
{\hsize}
{
l@{\extracolsep{0ptplus1fil}}
c@{\extracolsep{0ptplus1fil}}
c@{\extracolsep{0ptplus1fil}}
c@{\extracolsep{0ptplus1fil}}
c@{\extracolsep{0ptplus1fil}}}\hline
${\mathpzc f}$ & ${\mathpzc l}$ & ${\mathpzc s}$ & $\Delta{\mathpzc u}$ & $\Delta{\mathpzc s}$ \\\hline
${\mathpzc n}_{\mathpzc f}$  & $\phantom{-}3277\ $  & $\phantom{-}3446\ $  & $\phantom{-}1114\ $ & $-562.7\ $  \\
$a_1^{\mathpzc f}$  & $\phantom{-}3001\ $ & $\phantom{-}2242\ $ & $\phantom{-}2220\ $ & $\phantom{-}3747\ $  \\
$a_2^{\mathpzc f}$  & $-1820\ $  & $-2022\ $ & $-2145\ $ & $-994.3\ $   \\
$a_3^{\mathpzc f}$  & $-1464$ & $-1016\ $ & $-1193\ $ & $-1032\ $   \\
$a_4^{\mathpzc f}$  & $\phantom{-}31.33\ $ & $\phantom{-}229.9\ $ & $\phantom{-}367.8\ $ & $-70.84\ $   \\
$a_5^{\mathpzc f}$  & $\phantom{-}403.9\ $ & $\phantom{-}267.1\ $ & $\phantom{-}429.4\ $ & $\phantom{-}63.21\ $   \\
$a_6^{\mathpzc f}$  & $\phantom{-}117.3\ $ & $\phantom{-}24.94\ $ & $-30.81\ $ & $-31.14\ $  \\
$a_7^{\mathpzc f}$  & $-65.04\ $ & $-43.41\ $  & $-138.1\ $ & $-16.47\ $  \\
$a_8^{\mathpzc f}$  & $-42.65\ $ & $-17.42\ $ & $-26.72\ $ & $\phantom{-}3.160\ $  \\
$a_9^{\mathpzc f}$  & $\phantom{-}4.021\ $ & $\phantom{-}3.147\ $ &$\phantom{-}23.73\ $  & $\phantom{-}0.267\ $   \\
$a_{10}^{\mathpzc f}$  & $\phantom{-}6.668\ $ & $\phantom{-}3.061\ $ & $\phantom{-}9.747\ $ & $-0.904\ $ \\\hline
\end{tabular*}
\end{table}


\end{document}